%% file: ms.tex
\documentclass[twocolumn,resetfootnote]{aastex701}

\usepackage[figuresright]{rotating} 
\usepackage{array}

\usepackage{lipsum}  

\usepackage{amsmath}

\newcolumntype{L}[1]{>{\raggedright\let\newline\\\arraybackslash\hspace{0pt}}m{#1}}
\newcolumntype{C}[1]{>{\centering\let\newline\\\arraybackslash\hspace{0pt}}m{#1}}
\newcolumntype{R}[1]{>{\raggedleft\let\newline\\\arraybackslash\hspace{0pt}}m{#1}}

\newcommand{\A}{\text{\normalfont\AA}}

\def \hbeta{H$\beta$}

\def \halpha{H$\alpha$}

\def \h2{{\rm H_{2}}}

\def \hbeta{H$\beta$}
\def \halpha{H$\alpha$}
\def \cplus{[\ion{C}{2}]}

\def \oii{[\ion{O}{2}]}
\def \oiii{[\ion{O}{3}]}
\def \nii{[\ion{N}{2}]}

\def \dn4000{D_{{\rm n}}(4000) }

\def \x{$\times$}

\def \Rthree{([\ion{O}{3}]$_{\rm 5007}$ + [\ion{O}{3}]$_{\rm 4959}$)/H$\beta$}
\def \Rtwo{[\ion{O}{2}]$_{\rm 3727}$/H$\beta$}

\def \SR{[\ion{S}{2}]$_{\rm 6732}$/[\ion{S}{2}]$_{\rm 6718}$}
\def \Toii{$T_e({\rm O^{+} })$}
\def \Toiii{$T_e({\rm O^{++} })$}

\graphicspath{{./}{figures/}}

\begin{document}

\title{The ALPINE-CRISTAL-JWST Survey:\\
The Fast Metal Enrichment of Massive Galaxies at $z\sim5$}
\suppressAffiliations

\author[0000-0002-9382-9832]{Andreas L. Faisst}
\affiliation{IPAC, California Institute of Technology, 1200 E. California Blvd. Pasadena, CA 91125, USA}
\email{afaisst@caltech.edu}
\correspondingauthor{Andreas L. Faisst}


\author[0009-0004-1270-2373]{Lun-Jun Liu}
\affiliation{Physics Department, California Institute of Technology, 1200 E California Blvd, Pasadena, CA 91125, USA}
\email{}

\author[0000-0003-0225-6387]{Yohan Dubois}
\affiliation{Institut d’Astrophysique de Paris, UMR 7095, CNRS, Sorbonne Universit\'e, 98 bis boulevard Arago, 75014 Paris, France}
\email{}

\author[0000-0002-0984-7713]{Yurina Nakazato}
\affiliation{Center for Computational Astrophysics, Flatiron Institute, 162 5th Avenue, New York, NY 10010, USA}
\affiliation{Department of Physics, The University of Tokyo, 7-3-1 Hongo, Bunkyo, Tokyo 113-0033, Japan}
\email{}

\author[0009-0001-5733-8307]{Omima Osman}
\affiliation{INAF - Astronomical Observatory of Trieste, via G.B. Tiepolo 11, I-34143 Trieste, Italy}
\affiliation{University of Khartoum - Department of Physics, Al-Gamaa Ave, 11115 Khartoum, Sudan}
\email{}

\author[0000-0002-7129-5761]{Andrea Pallottini}
\affiliation{Dipartimento di Fisica ``Enrico Fermi'', Universit\'{a} di Pisa, Largo Bruno Pontecorvo 3, Pisa I-56127, Italy}
\email{}

\author[0000-0002-3258-3672]{Livia Vallini}
\affiliation{INAF – Osservatorio di Astrofisica e Scienza dello Spazio di Bologna, Via Gobetti 93/3, 40129 Bologna, Italy}
\email{}


\author[0000-0001-7201-5066]{Seiji Fujimoto}
\affiliation{Department of Astronomy, The University of Texas at Austin, Austin, TX 78712, USA}
\affiliation{David A. Dunlap Department of Astronomy and Astrophysics, University of Toronto, 50 St. George Street, Toronto, Ontario, M5S 3H4, Canada}
\email{}

\author[0000-0001-5846-4404]{Bahram Mobasher}
\affiliation{Department of Physics and Astronomy, University of California, Riverside, 900 University Avenue, Riverside, CA 92521, USA}
\email{}

\author[0000-0002-7964-6749]{Wuji Wang}
\affiliation{IPAC, California Institute of Technology, 1200 E. California Blvd. Pasadena, CA 91125, USA}
\email{}

\author[0000-0001-8792-3091]{Yu-Heng Lin}
\affiliation{IPAC, California Institute of Technology, 1200 E. California Blvd. Pasadena, CA 91125, USA}
\email{}


\author[0000-0001-5758-1000]{Ricardo O. Amor\'{i}n}
\affiliation{Instituto de Investigaci\`on Multidisciplinar en Ciencia y Tecnología, Universidad de La Serena, Ra\'ul Bitr\'an 1305, La Serena, Chile}
\affiliation{Departamento de Astronom\'ia, Universidad de La Serena, Av. Juan Cisternas 1200 Norte, La Serena, Chile}
\email{}

\author[0000-0002-6290-3198]{Manuel Aravena}
\affiliation{Instituto de Estudios Astrof\'isicos, Facultad de Ingenier\'ia y Ciencias, Universidad Diego Portales, Av. Ej\'ercito Libertador 441, Santiago 8370191, Chile}
\affiliation{Millenium Nucleus for Galaxies (MINGAL), Av. Ej\'ercito 441, Santiago 8370191, Chile}
\email{}

\author[0000-0002-9508-3667]{R. J. Assef}
\affiliation{Instituto de Estudios Astrof\'isicos, Facultad de Ingenier\'ia y Ciencias, Universidad Diego Portales, Av. Ej\'ercito Libertador 441, Santiago 8370191, Chile}
\email{}

\author[0000-0003-4569-2285]{Andrew J. Battisti}
\affiliation{International Centre for Radio Astronomy Research (ICRAR), The University of Western Australia, M468, 35 Stirling Highway, Crawley, WA 6009, Australia}
\affiliation{Research School of Astronomy and Astrophysics, Australian National University, Cotter Road, Weston Creek, ACT 2611, Australia}
\affiliation{ARC Center of Excellence for All Sky Astrophysics in 3 Dimensions (ASTRO 3D), Australia}
\email{}

\author[0000-0002-3915-2015]{Matthieu B\'ethermin}
\affiliation{Universit\'e de Strasbourg, CNRS, Observatoire astronomique de Strasbourg, UMR 7550, 67000 Strasbourg, France}
\email{}

\author[0000-0003-0946-6176]{M\'ed\'eric Boquien}
\affil{Universit\'e C\^ote d'Azur, Observatoire de la C\^ote d'Azur, CNRS, Laboratoire Lagrange, 06000, Nice, France}
\email{}

\author[0000-0002-6716-4400]{Paolo Cassata}
\affiliation{Dipartimento di Fisica e Astronomia Galileo Galilei Universit{\`a} degli Studi di Padova, Vicolo dell’Osservatorio 3, 35122 Padova Italy}
\affiliation{INAF - Osservatorio Astronomico di Padova, Vicolo dell’Osservatorio 5, I-35122, Padova, Italy}
\email{}

\author[0000-0001-9759-4797]{Elisabete da Cunha}
\affiliation{International Centre for Radio Astronomy Research (ICRAR), The University of Western Australia, M468, 35 Stirling Highway, Crawley, WA 6009, Australia}
\affiliation{ARC Center of Excellence for All Sky Astrophysics in 3 Dimensions (ASTRO 3D), Australia}
\email{}

\author[0009-0007-7842-9930]{Poulomi Dam}
\affiliation{Dipartimento di Fisica e Astronomia Galileo Galilei Universit{\`a} degli Studi di Padova, Vicolo dell’Osservatorio 3, 35122 Padova Italy}
\email{}

\author[0000-0002-6220-9104]{Gabriella de Lucia}
\affiliation{INAF - Astronomical Observatory of Trieste, via G.B. Tiepolo 11, I-34143 Trieste, Italy}
\affiliation{IFPU - Institute for Fundamental Physics of the Universe, via Beirut 2, 34151, Trieste, Italy}
\email{}

\author[0000-0001-9419-6355]{Ilse De Looze}
\affiliation{Sterrenkundig Observatorium, Ghent University, Krijgslaan 281 - S9, B-9000 Gent, Belgium}
\email{}

\author[0000-0003-0348-2917]{Miroslava Dessauges-Zavadsky}
\affiliation{Observatoire de Gen\`eve, Universit\'e de Gen\`eve, Chemin Pegasi 51, 1290 Versoix, Switzerland}
\email{}

\author[0000-0002-9400-7312]{Andrea Ferrara}
\affiliation{Scuola Normale Superiore, Piazza dei Cavalieri 7, I-56126 Pisa, Italy}
\email{}

\author[0000-0002-4462-0709]{Kyle Finner}
\affiliation{IPAC, California Institute of Technology, 1200 E. California Blvd. Pasadena, CA 91125, USA}
\email{}

\author[0000-0003-4744-0188]{Fabio Fontanot}
\affiliation{INAF - Astronomical Observatory of Trieste, via G.B. Tiepolo 11, I-34143 Trieste, Italy}
\affiliation{IFPU - Institute for Fundamental Physics of the Universe, via Beirut 2, 34151, Trieste, Italy}
\email{}

\author[0000-0002-9122-1700]{Michele Ginolfi}
\affiliation{Universit\`a di Firenze, Dipartimento di Fisica e Astronomia, via G. Sansone 1, 50019 Sesto Fiorentino, Florence, Italy}
\affiliation{INAF --- Arcetri Astrophysical Observatory, Largo E. Fermi 5, I-50125, Florence, Italy}
\email{}

\author[0009-0008-1835-7557]{Diego A. G\'{o}mez-Espinoza}
\affiliation{Instituto de F\'{i}sica y Astronom\'{i}a, Universidad de Valpara\'{i}so, Avda. Gran Breta\~{n}a 1111, Valpara\'{i}so, Chile}
\email{}

\author[0000-0002-5836-4056]{Carlotta Gruppioni}
\affiliation{INAF – Osservatorio di Astrofisica e Scienza dello Spazio di Bologna, Via Gobetti 93/3, 40129 Bologna, Italy}
\email{}

\author[0009-0005-8932-7783]{Nicol Guti\'errez-Vera}
\affiliation{Departamento de Astronom{\'i}a, Universidad de Concepci{\'o}n, Barrio Universitario, Concepci{\'o}n, Chile}
\affiliation{Millenium Nucleus for Galaxies (MINGAL), Av. Ej\'ercito 441, Santiago 8370191, Chile}
\email{}

\author[0009-0003-3097-6733]{Ali Hadi}
\affiliation{Department of Physics and Astronomy, University of California, Riverside, 900 University Avenue, Riverside, CA 92521, USA}
\email{}

\author[0000-0002-2775-0595]{Rodrigo Herrera-Camus}
\affiliation{Departamento de Astronom{\'i}a, Universidad de Concepci{\'o}n, Barrio Universitario, Concepci{\'o}n, Chile}
\affiliation{Millenium Nucleus for Galaxies (MINGAL), Av. Ej\'ercito 441, Santiago 8370191, Chile}
\email{}

\author[0000-0002-3301-3321]{Michaela Hirschmann}
\affiliation{Institute of Physics, Laboratory for galaxy evolution, EPFL, Observatoire de Sauverny, Chemin Pegasi 51, 1290 Versoix, Switzerland}
\email{}

\author[0009-0008-9801-2224]{Eduardo Ibar}
\affiliation{Instituto de F\'{i}sica y Astronom\'{i}a, Universidad de Valpara\'{i}so, Avda. Gran Breta\~{n}a 1111, Valpara\'{i}so, Chile}
\affiliation{Millenium Nucleus for Galaxies (MINGAL), Av. Ej\'ercito 441, Santiago 8370191, Chile}
\email{}

\author[0000-0003-4268-0393]{Hanae Inami}
\affiliation{Hiroshima Astrophysical Science Center, Hiroshima University, 1-3-1 Kagamiyama, Higashi-Hiroshima, Hiroshima 739-8526, Japan}
\email{}

\author[0000-0001-9187-3605]{Jeyhan S. Kartaltepe}
\email{jeyhan@astro.rit.edu}
\affiliation{Laboratory for Multiwavelength Astrophysics, School of Physics and Astronomy, Rochester Institute of Technology, 84 Lomb Memorial Drive, Rochester, NY 14623, USA}

\author[0000-0002-6610-2048]{Anton M. Koekemoer}
\affiliation{Space Telescope Science Institute, 3700 San Martin Dr., Baltimore, MD 21218, USA} 
\email{}

\author[0000-0003-1041-7865]{Mahsa Kohandel}
\affiliation{Scuola Normale Superiore, Piazza dei Cavalieri 7, I-56126 Pisa, Italy}
\email{}

\author[0000-0001-7457-4371]{Lilian L. Lee}
\affiliation{Max-Planck-Institute f\"ur extratarrestrische Physik, Giessenbachstrasse 1, 85748 Garching, Germany}
\email{}

\author{Yuan Li}
\affiliation{Department of Physics and Astronomy and George P. and Cynthia Woods Mitchell Institute for Fundamental Physics and Astronomy, Texas A\&M University, 4242}
\email{}

\author[0000-0002-8136-8127]{Juan Molina}
\affiliation{Instituto de F\'{i}sica y Astronom\'{i}a, Universidad de Valpara\'{i}so, Avda. Gran Breta\~{n}a 1111, Valpara\'{i}so, Chile}
\affiliation{Millenium Nucleus for Galaxies (MINGAL), Av. Ej\'ercito 441, Santiago 8370191, Chile}
\email{}

\author[0000-0001-6652-1069]{Ambra Nanni}
\affiliation{National Centre for Nuclear Research, ul. Pasteura 7, 02-093 Warsaw, Poland}
\affiliation{INAF - Osservatorio astronomico d'Abruzzo, Via Maggini SNC, 64100, Teramo, Italy}
\email{}

\author[0000-0002-7064-4309]{Desika Narayanan}
\affiliation{Department of Astronomy, University of Florida, 211 Bryant Space Sciences Center, Gainesville, FL 32611, USA}
\affiliation{Cosmic Dawn Center (DAWN), Copenhagen, Denmark}
\email{}

\author[0000-0002-7412-647X]{Francesca Pozzi}
\affiliation{Dipartimento di Fisica e Astronomia ``Augusto Righi'', Alma Mater Studiorum, Universi\'a di Bologna, Via Gobetti 93/2, 40129 Bologna, Italy}
\affiliation{INAF – Osservatorio di Astrofisica e Scienza dello Spazio di Bologna, Via Gobetti 93/3, 40129 Bologna, Italy}
\email{}

\author[0000-0003-1682-1148]{Monica Relano}
\affiliation{Dept. Física Te\'{o}rica y del Cosmos, Campus de Fuentenueva, Edificio Mecenas, Universidad de Granada, E-18071, Granada, Spain}
\affiliation{Instituto Universitario Carlos I de Física Te\'{o}rica y Computacional, Universidad de Granada, 18071, Granada, Spain}
\email{}

\author[0000-0002-9948-3916]{Michael Romano}
\affiliation{Max-Planck-Institut für Radioastronomie, Auf dem H\"ugel 69, 53121 Bonn, Germany}
\affiliation{INAF - Osservatorio Astronomico di Padova, Vicolo dell’Osservatorio 5, I-35122, Padova, Italy}
\email{}

\author[0000-0002-1233-9998]{David B. Sanders}
\affiliation{Institute for Astronomy, University of Hawaii, 2680 Woodlawn Drive, Honolulu, HI 96822, USA}
\email{}

\author[0000-0002-0000-6977]{John D. Silverman}
\affiliation{Kavli Institute for the Physics and Mathematics of the Universe (Kavli IPMU, WPI), UTIAS, Tokyo Institutes for Advanced Study, University of Tokyo, Chiba, 277-8583, Japan}
\affiliation{Department of Astronomy, Graduate School of Science, The University of Tokyo, 7-3-1 Hongo, Bunkyo, Tokyo 113-0033, Japan}
\affiliation{Center for Data-Driven Discovery, Kavli IPMU (WPI), UTIAS, The University of Tokyo, Kashiwa, Chiba 277-8583, Japan}
\affiliation{Center for Astrophysical Sciences, Department of Physics \& Astronomy, Johns Hopkins University, Baltimore, MD 21218, USA}
\email{}

\author[0000-0002-2906-2200]{Laura Sommovigo}
\affiliation{Center for Computational Astrophysics, Flatiron Institute, 162 5th Avenue, New York, NY 10010, USA}
\email{}

\author[0000-0003-3256-5615]{Justin Spilker}
\affiliation{Department of Physics and Astronomy and George P. and Cynthia Woods Mitchell Institute for Fundamental Physics and Astronomy, Texas A\&M University, 4242}
\email{}

\author[0000-0002-0498-5041]{Akiyoshi Tsujita}
\affiliation{Institute of Astronomy, Graduate School of Science, The University of Tokyo, 2-21-1 Osawa, Mitaka, Tokyo 181-0015, Japan}
\email{}

\author[0000-0003-4891-0794]{Hannah \"Ubler}
\affiliation{Max-Planck-Institute f\"ur extratarrestrische Physik, Giessenbachstrasse 1, 85748 Garching, Germany}
\email{}

\author[0000-0002-2645-679X]{Keerthi Vasan G.C.}
\affiliation{The Observatories of the Carnegie Institution for Science, 813 Santa Barbara Street, Pasadena, CA 91101, USA}
\email{}

\author[0009-0007-1304-7771]{Enrico Veraldi}
\affiliation{Scuola Internazionale Superiore Studi Avanzati (SISSA), Physics Area, Via Bonomea 265, 34136 Trieste, Italy}
\email{}

\author[0000-0002-5877-379X]{Vincente Villanueva}
\affiliation{Departamento de Astronom{\'i}a, Universidad de Concepci{\'o}n, Barrio Universitario, Concepci{\'o}n, Chile}
\email{}

\author[0000-0003-3864-068X]{Lizhi Xie}
\affiliation{Tianjin Normal University, Binshuixidao 393, 300387, Tianjin, China}
\email{}

\author[0000-0002-2318-301X]{Gianni Zamorani}
\affiliation{INAF – Osservatorio di Astrofisica e Scienza dello Spazio di Bologna, Via Gobetti 93/3, 40129 Bologna, Italy}
\email{}

\collaboration{all}{(Affiliations can be found after the references)}





\begin{abstract}
We present the stellar mass-metallicity relation (MZR) and mass-metallicity-star formation relation (``fundamental metallicity relation''; FMR) of $18$ massive ($\rm \log(M_\star/M_\odot) = 9.5 - 11$) main-sequence galaxies at $z\sim5$ from the ALPINE-CRISTAL-JWST sample. This sample complements recent studies by JWST at up to two orders of magnitude lower stellar masses. The metallicities are derived using strong optical lines, and verified by temperature-based oxygen abundance measurements for five galaxies for which faint auroral lines are detected.
We find that the metal abundance evolves, on average, from $40\%$ to $60\%$ solar between $z\sim5$ and cosmic noon ($z\sim2$) at the massive end of the MZR, suggesting already significant metal enrichment at early times.
The FMR at $z=5$ exhibits a $5\times$ larger scatter (preferentially to lower metallicities) compared the local FMR relation. This scatter can be explained by a bursty star formation and the direct build-up of metals in early galaxies as well as differences in age and outflow efficiencies.
Capitalizing on all available samples, we find that the observed MZR and FMR over three orders of stellar mass is generally in good agreement with results from cosmological simulation, although some underestimate the metal enrichment at low stellar masses. This may be due to too efficient metal-rich outflows. We show that the ALPINE-CRISTAL-JWST galaxies likely joined the current FMR at $z\sim10$ and will evolve into massive ($\rm \log(M_\star/M_\odot) \sim 11.4$) galaxies with super-solar metallicities by $z=0$.


\end{abstract}



\section{Introduction} \label{sec:intro}

Metallicity, the relative abundance of heavy elements to hydrogen, is a fundamental observable in studying the evolution of galaxies \citep{tinsley80,pagel97,maiolino19}. Its role, to constrain the mass assembly of galaxies and the feedback process due to infall and outflow of gas, is widely used to study the chemical enrichment of galaxies. 
The stellar mass vs. metallicity relation (MZR) of galaxies observed in the local universe shows the tight correlation of gas-phase metallicity increasing with the stellar mass \citep{tremonti04,mannucci10,berg12,andrews13,sanchez17,curti20}. Advances in near-infrared (near-IR) spectroscopy have allowed measurements of rest-frame optical line ratio diagnostics of the metallicity of galaxies at higher redshifts. Using statistically large samples, independent studies have shown that, in general, the metal abundance at a given stellar mass decreases with increasing redshift \citep{maiolino08,kashino17,maiolino19,sanders21,papovich22,strom22}. The MOSDEF \citep{kriek15} and KBSS \citep{rudie12,steidel14} surveys performed extensive observations of the MZR to $z\sim3$, confirming the evolution of this relation with look-back time \citep{sanders18,strom17}. The evolution and scatter in the mass-metallicity relation with redshift has strong implications towards the formation and subsequent metal enrichment of galaxies and the process of stellar evolution at earlier times.

Observations by the James Webb Space Telescope (JWST) have now given access to rest-frame optical diagnostic lines at the reionization epoch ($z\sim5-7$), allowing the observation of the MZR to the highest redshifts \citep{nakajima23,curti24,marszewski24,morishita24,rowland25,sarkar25}. This enables the study of the metal enrichment process undergone by pristine gas and the infall process in the inter-stellar medium (ISM) of early galaxies.

A serious challenge for such studies was the lack of reliable calibration for metallicity measurements at high redshifts. Previous studies assumed local calibration with no significant evolution with cosmic time. This assumption directly affect the reliability of the mass-metallicity relation at $z > 3$. 
In a recent study, \citet{sanders24} performed a calibration between rest-frame optical line ratios and metallicities in the redshift range $z\sim2-9$ using a method based on the electron temperature ($T_e$) measurement. This technique utilizes optical emission line ratios involving faint auroral lines, using transitions resulting from different upper energy levels (e.g., [\ion{O}{3}]$_{4363}$/[\ion{O}{3}]$_{5007}$ or [\ion{O}{2}]$_{7322,7332}$/[\ion{O}{2}]$_{3727}$ ratios). The estimated $T_e$ can then be used to convert dust-corrected flux ratios of ionic Oxygen lines to Hydrogen recombination lines (e.g., \oiii/\hbeta~and \oii/\hbeta)\footnote{In the following, we define \oiii~$\equiv$~[\ion{O}{3}]$_{5007}$ and \oii~$\equiv$~[\ion{O}{2}]$_{3727+3730}$.} into O/H abundance ratios to construct metallicity calibration by fitting functional forms to the relation between different line ratios and O/H \citep{pettini04,nagao06,maiolino08,curti20}. This calibration procedure needs measurement of various rest-frame optical lines -- [\ion{O}{3}]$_{4363}$, \oiii, \oii, \hbeta, [\ion{O}{2}]$_{7322,7332}$~-- of galaxies at high redshifts. Before JWST, it has therefore been difficult to implement this technique at high redshifts due to inaccessibility of rest-frame optical lines and the faintness of such auroral lines.
Now, all of these lines are accessible up to $z\sim6$ simultaneously, hence allowing a consistent metallicity measurement across the past 12 billion years.

Extending the dimensionality of the MZR by the star formation rate (SFR) led to the recognition of the fundamental metallicity relation (FMR) at low redshifts \citep[e.g.,][]{ellison08,mannucci10} which was soon explained by (semi-) analytical models \citep[e.g.,][]{dave12,dayal13,lilly13,feldmann15,delucia20}. While this relation is now well established out to $z\sim3$ \citep[e.g.,][]{perezmontero13,wuyts14,yabe14,zahid14,zahid14a,salim15,sanders15,guo16,onodera16,kashino17,cullen19,sanders21,henry21,topping21}, although some works suggest no evidence of an FMR at $z=2-3$ \citep[e.g.,][]{steidel14,korhonen-cuestas25}, only recently with JWST has it been measured at higher redshifts. These data, importantly only focusing on low-mass galaxies, suggest an evolution of the FMR compared to the local relation \citep[e.g.,][]{curti24,robertsborsani24,pollock26,scholte25,sarkar25,stanton26}.

\input{table2.tex}

Current JWST observations referenced above have derived the MZR at stellar masses of $\rm \log(M_\star/M_\odot) \lesssim 9.5$ at high redshifts.
The relation (and its evolution) for more massive galaxies in the early Universe are still unknown. This is mainly because such galaxies are rare and not covered by deep pencil-beam surveys (for example a $5\times10^{10}\,{\rm M_\odot}$ galaxy is ten times rarer than a $10^9\,{\rm M_\odot}$ galaxy, \citealt{shuntov25a}).
Similarly, the question of a possible evolution of the FMR across the full mass range from $10^{7}-10^{11}\,{\rm M_\odot}$ at high redshifts is not answered.
Tracing the MZR and FMR, thus the metal enrichment, in massive galaxies is an important tracer of galaxy growth and, in relation to lower stellar masses, holds valuable information on the overall evolution of galaxies.

In this work, we present constraints on the MZR and FMR at $z\sim5$ from galaxies of the ALPINE-CRISTAL-JWST survey  \citep{faisst26,fujimoto25}. This survey provides JWST/NIRSpec IFU observations for a sample of $18$ main-sequence galaxies at the end of the epoch of reionization (EoR) at $z=4-6$. Notably, all galaxies also have ALMA observations of \cplus$_{158\,{\rm \mu m}}$ and far-IR continuum at comparable resolution to JWST spectroscopy to jointly study the stars, gas, and dust at this epoch. The galaxies are massive at $\rm \log(M_\star/M_\odot) > 9.5$, thus complement the samples at lower stellar masses observed by other JWST surveys. We investigate their scatter on the MZR and their location on the FMR in relation to theoretical models to characterize the fast build-up of metals at these early times.

This paper is organized as follows: In Section~\ref{sec:measurements}, we describe the sample and outline the measurement of the metal abundances (with more details in Appendix~\ref{app:metal}). In Section~\ref{sec:results}, we present the observed MZR and FMR at $z\sim5$, which we then study in the light of a simple analytical model and cosmological simulations in Section~\ref{sec:discussion}. 
Throughout this work, we assume a $\Lambda$CDM cosmology with $H_0 = 70\,{\rm km\,s^{-1}\,Mpc^{-1}}$, $\Omega_\Lambda = 0.7$, and $\Omega_{\rm m} = 0.3$ and magnitudes are given in the AB system \citep{oke74}. We use a Chabrier initial mass function \citep[IMF;][]{chabrier03} calibration for stellar masses and SFRs.

\section{Data \& Measurements} \label{sec:measurements}

\subsection{The ALPINE-CRISTAL-JWST Sample}
This work is based on the $18$ of the ALPINE-CRISTAL-JWST survey. The survey, including science goals and basic measurements, is described in its entirety in \citet{faisst26}. The JWST/NIRSpec IFU data reduction and spatially resolved metallicity measurements are described in \citet{fujimoto25}. Here, we give a brief overview of the sample and the survey.

The ALPINE-CRISTAL-JWST survey targeted $18$ main-sequence galaxies at $4.4 < z < 5.9$ ($\left<z\right>\sim5$) from the ALMA-ALPINE program \citep[\cplus$_{158\,{\rm \mu m}}$~and $158\,{\rm \mu m}$ dust continuum observations at $0.7\arcsec$ spatial resolution;][]{lefevre20,bethermin20,faisst20b} at $\rm \log(M_\star/M_\odot) \sim 9.5-11.0$ and $\rm SFR \sim 10-700\,{\rm M_\odot\,yr^{-1}}$ with rest-frame optical JWST/NIRSpec IFU observations (\#3045; PI: Faisst) at $0.15\arcsec$ spatial resolution in G235M/F170LP and G395M/F290LP ($R\sim 1000$). As part of the selection process, all galaxies have HST rest-frame ultra-violet (UV) imaging, JWST/NIRCam rest-frame optical imaging \citep[COSMOS-Web program;][]{casey22}, and additional higher resolution ALMA observations of the \cplus$_{158\,{\rm \mu m}}$ line and the $158\,{\rm \mu m}$ dust continuum at $0.3\arcsec$ spatial resolution from the ALMA-CRISTAL survey \citep{herreracamus25}. All galaxies reside in the COSMOS field \citep{scoville07,koekemoer07}, which provides a wealth of ancillary data.

Although none of our targets can be unambiguosly defined as AGN based on the BTP diagram \citep{baldwin81}, one of them ({\it DC-873756}) shows strong \nii~emission, making it a potential BPT candidate AGN. In addition, other six targets ({\it DC-417567}, {\it DC-519281}, {\it DC-536534}, {\it DC-683613}, {\it DC-848185}, and {\it DC-873321}) are potential type 1 AGN due to broad ($>500\,{\rm km/s}$) \halpha~emission \citep{ren25}.

Some physical parameters are summarized in Table~\ref{tab:physical} \citep[more can be found in][]{faisst26}.
The stellar masses adopted in this work are derived in \citet{mitsuhashi24} using the \texttt{CIGALE} code \citep{boquien19,burgarella05}. The SFRs are derived from the total UV$+$far-IR emission \citep{faisst26}. The galaxy ages (light weighted) are taken from \citet{faisst20b} and are derived using \texttt{LePhare} \citep{arnouts99,ilbert06}. Four galaxies are not far-IR detected and we use the SED-fitting results from \citet{mitsuhashi24} to measure their SFR. \halpha-based SFRs extinction-corrected with Balmer decrement are derived using the \citet{kennicutt98} relation (converted to a \citet{chabrier03} IMF).

The ALPINE-CRISTAL-JWST survey provides the best-studied $z\sim5$ main-sequence galaxy sample at $\log({\rm M/M_\odot}) = 9.5 - 11.0$ at $1-2\,{\rm kpc}$ resolution to date. It is a benchmark in the study of galaxy evolution in the post-EoR era --- the connection between primordial galaxy formation in the EoR and mature galaxy evolution at cosmic noon.

\subsection{Metallicity Measurement}

From the new JWST/NIRSpec spectroscopy, we measured spatially integrated oxygen abundances (``metallicity'', defined as $Z \equiv 12+\log({\rm O/H})$) of the $18$ galaxies using strong optical emission lines with applied $T_e$ calibration from \citet{sanders24}. These strong lines are detected for all individual galaxies.
The metallicity measurements are detailed in \citet{faisst26} and summarized in Table~\ref{tab:physical} and in the following we give an overview.
In Appendix~\ref{app:metal}, we additionally compare different strong-line metallicity calibrators as well as compare these measurement to direct metallicity measurements from fainter auroral lines for a subsample of five galaxies.

The measurement of the strong-line metallicities is based on the dust-corrected optical lines fluxes of \oii, \hbeta, and \oiii~to derive oxygen abundances \citep[see review on metallicity measurements in ][]{kewley19}. All emission lines are extracted from the spectral cubes in a mask defined by the region of the $3\sigma$ JWST/NIRCam rest-frame optical continuum flux ({\it i.e.} tracing the stellar mass). But note the results do not depend significantly on the exact spatial outline. We do not use \halpha~and \nii~in the case of contamination by an AGN \citep{ren25} and the blending of the two lines in the medium resolution spectra.
Dust correction was applied via the Balmer decrement (\halpha/\hbeta) assuming case B recombination. Due to the coverage with both the G235M and G395M grating, the same lines could be used for all galaxies in the sample. We use the metallicity parameterization and calibration by \citet{sanders24}, which we found to lead to consistent metallicities as the parameterization by \citet{maiolino08} and \citet{hirschmann23} (Figure~\ref{fig:metcomparison}, Appendix~\ref{app:metal}).

The strong-line metallicity measurements were also compared to direct $T_e$-based measurements derived from fainter auroral lines ([\ion{O}{3}]$_{4363}$ and [\ion{O}{3}]$_{7322,7332}$) for a subset of five galaxies. For these galaxies, we find consistent metallicity measurement from the strong-line method an the $T_e$ method (see Appendix~\ref{app:metal} for details).

As mentioned previously, six of the ALPINE-CRISTAL-JWST galaxies are potential type 1 AGN candidates. However, as they do not seem to exhibit unusual line ratios typical for BPT AGN, we do not think that their possible AGN activity has a significant effect on the metallicity measurement.

\begin{figure*}[t!]
\centering
\includegraphics[angle=0,width=0.595\textwidth]{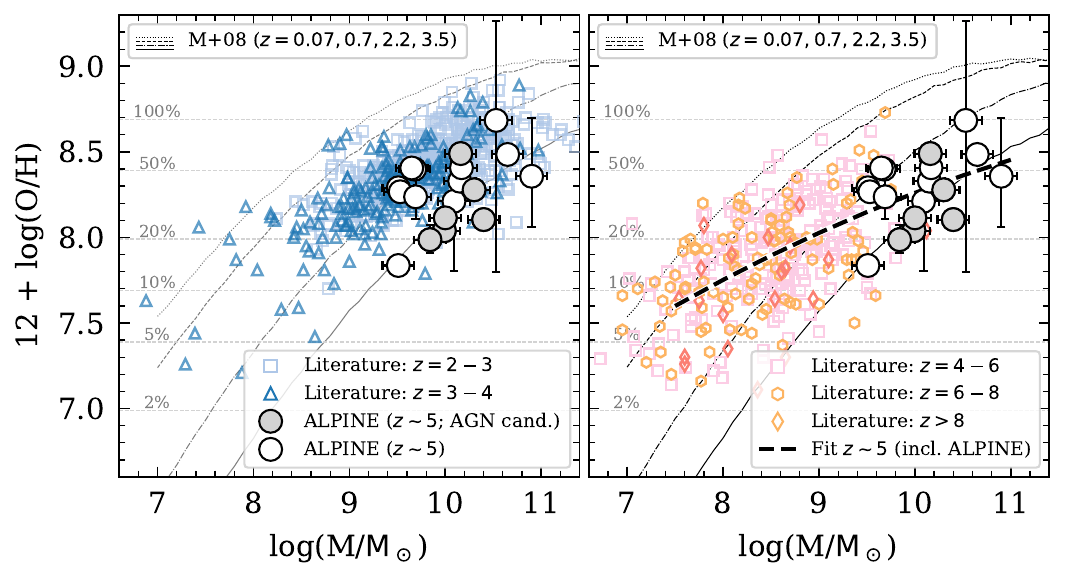}
\includegraphics[angle=0,width=0.395\textwidth]{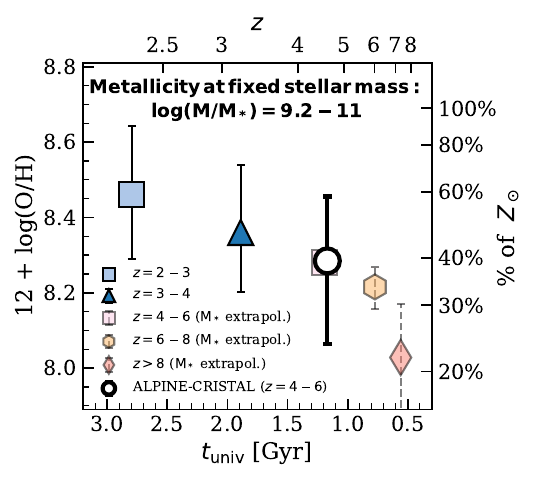}\vspace{-3mm}
\caption{
{\em Left and Middle:} Comparison of the MZR of the ALPINE-CRISTAL-JWST sample (open large circles; type 1 broad-line AGN candidates from \citet{ren25} are indicated in gray) to samples at $z<4$ ({\em left}) and $z>4$ ({\rm right}) from the literature \citep{steidel14,sanders15,nakajima23,morishita24,curti23,scholtz25,sanders24,stanton24,sanders25,sarkar25,rowland25,stanton26}. The relations from \citet{maiolino08} are indicated as lines as indicated in the legend. Metal abundances in percent of solar are shown as horizontal dashed lines. The best fit to the $z=4-6$ samples is indicated as dashed line ({\em middle} panel) with parameters listed in Table~\ref{tab:mzrfit}.
{\em Right:} Metal abundance evolution from cosmic noon to cosmic dawn at the fixed stellar mass range of the ALPINE-CRISTAL-JWST sample. Faint symbols with dashed errorbars (at $z>4$) are derived by extrapolating the MZR to the ALPINE-CRISTAL-JWST stellar mass range.
\label{fig:mzrobs}}
\end{figure*}

\section{The Mass-$Z$-SFR Relation of Massive Main-Sequence Galaxies at $z\sim5$}\label{sec:results}

The build-up of metals directly relates to the build-up of stellar mass over the lifetime of galaxies. However, second-order deviations from such a relation provide important insights into changes of metal production due to different processes such as star formation efficiency, outflows, inflows, and other physical events (e.g., supernovae or changes in the IMF).
With JWST, it has now become possible to constrain the metal content of galaxies close to the EoR at $z = 6$. So far, JWST provided such measurements for low-mass systems, generally well below $\rm \log(M_\star/{\rm M_\odot}) = 9.5$. The ALPINE-CRISTAL-JWST survey provides a sample of more massive $z\sim5$ galaxies that complements these other measurements and extends the relation between stellar mass, SFR, and metallicity to the massive end to $\rm \log(M_\star/{\rm M_\odot}) = 11$ at $z\sim5$.

\subsection{The Mass-Metallicity Relation (MZR)}

Figure~\ref{fig:mzrobs} shows the resulting MZR for the ALPINE-CRISTAL-JWST sample (large circles and type 1 AGN candidates indicated in gray) as well as a comparison to the MZR at lower redshifts.
The measured abundances of our sample range from $7.7-8.5$ in $\rm 12+\log(O/H)$, corresponding to $10-70\%$ of solar metallicity\footnote{We assume here a solar value of $12+\log({\rm O/H}) = 8.69$ \citep{allendeprieto01}, however, note that this value could be higher \citep{bergemann21}.}.
The left panel of Figure~\ref{fig:mzrobs} show the comparison of the ALPINE-CRISTAL-JWST sample to samples at $z=2-4$ from various literature \citep{maiolino08,steidel14,sanders15,morishita24,nakajima23,curti23,scholtz25,sanders24,sanders25,stanton24,stanton26}.
The middle panel shows the comparison to similar and higher redshifts (out to $z\sim10$) from similar literature and additionally including \citet{sarkar25} and the REBELS-IFU survey \citep{rowland25}.
These measurements use calibrations consistent with our work and we therefore do not expect significant systematics.
A comparison to cosmological simulations and analytical models is be discussed in Section~\ref{sec:discussioncosmological}.

The ALPINE-CRISTAL-JWST galaxies lie at the high-mass end of the distribution of JWST-measured metal abundances, thus ideally complementing existing JWST studies at lower masses. Their metal abundances compare reasonably well in extrapolation with the abundances of these lower-mass galaxies; they are generally about $0.5\,{\rm dex}$ more metal enriched compared to galaxies at $\rm \log(M_\star/{\rm M_\odot}) < 9$.
In the right panel of Figure~\ref{fig:mzrobs} we show the evolution of metal abundance as a function of time (or redshift) at a {\it fixed} stellar mass comparable to the ALPINE-CRISTAL-JWST sample. We note that only samples at $z<4$ cover a similar stellar mass range. For the comparison to literature samples at $z>4$, we extrapolated their metal abundances to the stellar mass of the ALPINE-CRISTAL-JWST sample assuming a power-law fit (these measurements are distinguished by dashed error bars).
From this panel we see that the metal abundances at a fixed stellar mass increased on average from $40\%$ to $60\%$ of solar from $z=5$ to cosmic noon (about a $50\%$ increase over two billion years). 
In the previous only $500$ million years (from $z\sim8$ to $z=5$), a similar increase in metals at fixed stellar mass is observed, indicating a fast metal enrichment at earlier times.
The fast metal enrichment at early cosmic times is further discussed in Section~\ref{sec:discussion}.

The thick black dashed line in the middle panel of Figure~\ref{fig:mzrobs} shows a fit using the \texttt{scipy.optimize} {\it Python} package to all available data at $z=4-6$ and $\rm \log(M_\star/M_\odot) = 7.5 - 11.0$ including the ALPINE-CRISTAL-JWST sample:\footnote{We adopt here a common functional form as in \citet{maiolino08}.}
\begin{equation}\label{eq:mzr}
    12+\log({\rm O/H}) = B \, \left( \log(M_\star/{\rm M_\odot}) - \log(M_0) \right)^2 + A,
\end{equation}
with best-fit parameters given in Table~\ref{tab:mzrfit}.

\begin{figure*}[t!]
\centering
\includegraphics[angle=0,width=0.48\textwidth]{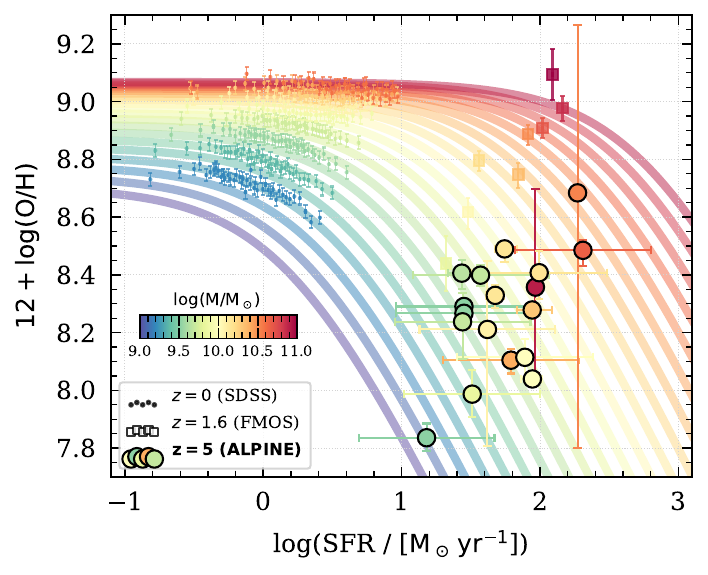}
\includegraphics[angle=0,width=0.48\textwidth]{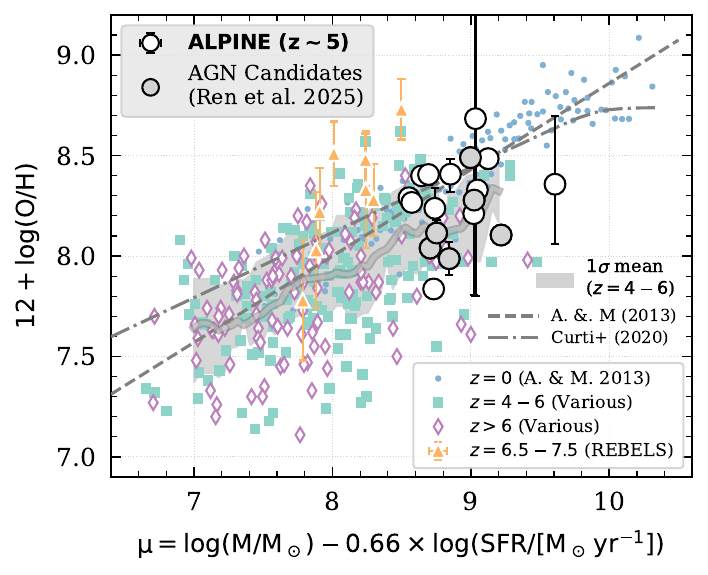}
\vspace{-3mm}
\caption{The stellar mass, metallicity, and SFR relation (``fundamental metallicity relation'', FMR).
{\it Left:} Data are shown at $z=0$ from SDSS (small circles), at $z\sim2$ from the FMOS survey \citep[large squares;][]{silverman15,kashino17}, and from the ALPINE-CRISTAL-JWST survey at $z\sim5$ (large circles; this work). The data are compared to an equilibrium gas-regulator model (see Equation~\ref{eq:fmr}) from \citet{lilly13} \citep[see also][]{maier14,onodera16}, calibrated by \citet{kashino17} to the SDSS $z=0$ measurements (colored lines). The symbols and lines are color-coded by stellar mass on the same scale.
{\it Right:} The collapsed FMR assuming $\mu_0=0.66$ \citep{andrews13}. The ALPINE-CRISTAL-JWST sample at $z\sim5$ is shown as large circles (type 1 AGN candidates from \citealt{ren25} indicated in gray). Other literature data are shown at $z=4-6$ and $z>6$ \citep{nakajima23,curti24,morishita24,sarkar25,heintz25} as well as from the REBELS-IFU survey \citep{rowland25}.
The small blue points are data at $z=0$ \citep{andrews13} with the corresponding fit shown as dashed gray line.
The dot-dashed gray line shows the $z=0$ parameterization by \citet{curti20}.
The gray band shows the $1\sigma$ percentiles of the literature and ALPINE-CRISTAL-JWST data at $z=4-6$.
\label{fig:fmr}}
\end{figure*}

\input{table1.tex}

\subsection{The Fundamental Metallicity Relation (FMR)}

In addition to the MZR it is known that galaxies also follow a similar relation between stellar mass, metallicity, and SFR \citep[the so-called ``fundamental metallicity relation'' or FMR, e.g.,][]{ellison08,laralopez10,mannucci10,andrews13,curti20}. This fundamental relation, specifically the anti-correlation between metallicity and SFR, has also been identified in analytical models and numerical simulations \citep{dave12,dayal13,lilly13,feldmann15,delucia20}.
They support the idea of reducing the metal abundances in the ISM due to the infall of pristine/metal-poor gas, which is related to an increased star formation.
This three-dimensional relation has been well studied out to $z=2-3$ \citep[e.g.,][]{steidel14,wuyts14,yabe14,zahid14,zahid14a,salim15,sanders15,guo16,onodera16,kashino17,cullen19,sanders21,henry21,topping21,korhonen-cuestas25}, however, no comprehensive studies at higher redshifts covering the full mass range of galaxies have been undertaken \citep[see recent compilation in][]{pollock26,stanton26}. Here we combine the massive ALPINE-CRISTAL-JWST sample with measurements at lower masses from the literature to test for the first time the universality of a FMR at $z\sim5$.

The left panel in Figure~\ref{fig:fmr} shows
that the $z\sim5$ data does roughly follow the expectations of an FMR, however with some scatter (discussed in Section~\ref{sec:fmrscatter}).
Specifically, this figure shows our new measurements from the ALPINE-CRISTAL-JWST sample at $z\sim5$ together with data points from SDSS at $z\sim0$ \citep{abazajian09,alam15} and FMOS at $z\sim 1.6$ \citep{silverman15,kashino17}. The lines show the common {\it equilibrium} gas-regulator (``bathtub'') models from \citet{lilly13} \citep[see also][]{maier14,dekel14,feldmann15,onodera16} calibrated by \citet{kashino17} to the $z=0$ data from SDSS. In this model, the equilibrium metallicity depends on stellar mass and SFR as follows:
\begin{equation}\label{eq:fmr}
\begin{split}
    Z_{\rm eq} & = Z_{\rm in} + \\
    & \frac{y}{1 + \lambda(1-R)^{-1} + \epsilon_{\rm out}^{-1}\left((1+\beta-b)\times\frac{{\rm SFR}}{M_\star} - 0.15\right) },
\end{split}
\end{equation} 
where $Z_{\rm in}$ is the metallicity of the inflowing gas, $y$ is the metal yield\footnote{Here, the yield is non-dimensional and defined as the mass of metals returned to the ISM per unit mass that is locked up in long-lived stars, {\it i.e.} $(1-R)$ times the mass of stars formed; see \citet{lilly13}.}, $R$ is the fraction of mass returned to the ISM, $\lambda=\lambda_0\times \left(\frac{M_\star}{10^{10}\,{\rm M_\odot}}\right)^a$ is the mass-loading factor, $\epsilon_{\rm out} = \epsilon_{\rm out,0}\times \left(\frac{M_\star}{10^{10}\,{\rm M_\odot}}\right)^b$ is the star formation efficiency (e.g., $1/{\rm Gyr}$), and $\beta = -0.25$ is the slope of the relation between stellar mass and specific SFR. For consistency, we assume here the same values as in \citet{kashino13}; $Z_{\rm in} = 0$, $\log(y) = 9.09$, $R=0.27$, $(\lambda_0,a) = (0.38,-0.61)$, and $(\epsilon_{\rm out,0},b) = (0.53,0.48)$. For the derivation of the subtraction $-0.15$, see equation~40 in \citet{lilly13}.
Both our analytical model and data points are color-coded by stellar mass in the same color scale.
The colors of the ALPINE-CRISTAL-JWST galaxies and the equilibrium model curves (notably calibrated to the $z=0$ sample) agree well, suggesting that galaxies at $z\sim5$ follow a similar FMR as local galaxies, however with a larger scatter as discussed in Section~\ref{sec:fmrscatter}.

The right panel in Figure~\ref{fig:fmr} visualizes the above in more detail by showing an angled slice through the FMR by defining $\mu \equiv \log({\rm M_\star}) - \mu_0\times\log({\rm SFR})$ where $\mu_0=0.66$ based on a fit to galaxies at $z=0$ \citep{andrews13}.
The fundamentality in the FMR comes from the fact that $\mu_0$ minimizes the scatter of the FMR with respect to the MZR ({\em i.e.} $\mu_0 = 0$).
The ALPINE-CRISTAL-JWST galaxies are shown as large symbols together with literature data in two redshift bins at $z=4-6$ and $z>6$ \citep{nakajima23,curti24,morishita24,sarkar25,heintz25,rowland25}. Note that we find a tentative trend of type 1 AGN candidate host galaxies to show a lower metallicity, however this claim needs to be confirmed with larger samples.

Of interest is the comparison to the local FMR parameterized as ${\rm 12+log(O/H)} = 0.43\,\mu + 4.56$ by \citet{andrews13} (dashed line, right panel of Figure~\ref{fig:fmr}). 
To first order galaxies at $z=4-6$ follow this local relation well but to second order the data suggest a scatter and offset towards lower metallicity at a given $\mu$. This indicates an evolving FMR compared to $z=0$.
Although in broad agreement, we note that the recent study by \citet{curti24} finds a larger offset and scatter, hence a stronger evolution of the FMR. This is due to a different calibration of the local FMR --- indeed, the evolution is stronger if the local parameterization by \citet{curti20} (dot-dashed line)\footnote{Note that \citet{curti20} derives a $\mu_0=0.65$ for the high sSFR SDSS sample, which is in agreement with the value adopted here ($0.66$) derived by \citet{andrews13}. For consistency, we use $0.66$ in this work.} is used.
However, the result at high stellar masses (also showing a downward scatter in the FMR) is independent on the assumption of the $z=0$ FMR relation.

Finally, the downward scatter relative to the $z=0$ FMR relation suggests a negative offset in the FMR towards the early universe. A similar trend was also found by other literature \citep{curti24,stanton26,pollock26}. However, we point out that the data also suggest an asymmetric scatter to the mean FMR relation at $z=5$, which is further discussed in Section~\ref{sec:bursty}.

\section{Discussion}\label{sec:discussion}

In the previous section, we have reported on the MZR (Figure~\ref{fig:mzrobs}) as well as the FMR (Figure~\ref{fig:fmr}) at $z\sim5$ for massive galaxies in the ALPINE-CRISTAL-JWST sample. The data show that there is only a slight evolution in the metal abundance from the EoR to cosmic noon at a given stellar mass (Figure~\ref{fig:mzrobs}) and in addition galaxies at $z\sim5$ lie to first order on a similar FMR as galaxies at $z=0$ and $z\sim1.6$ (Figure~\ref{fig:fmr}). This indicates a fast build-up of metals at early times and subsequent steady main-sequence enrichment to the local universe.
However, to second order there are differences to the average low-redshift galaxy population: there is a significant scatter (and offset) in the MZR and FMR specifically towards low metallicities at a given stellar mass compared to the local relations.
In the following, we investigate the MZR and FMR scatter and compare the observations to a simple analytical model and cosmological simulations.

\begin{figure}[t!]
\centering
\includegraphics[angle=0,width=\columnwidth]{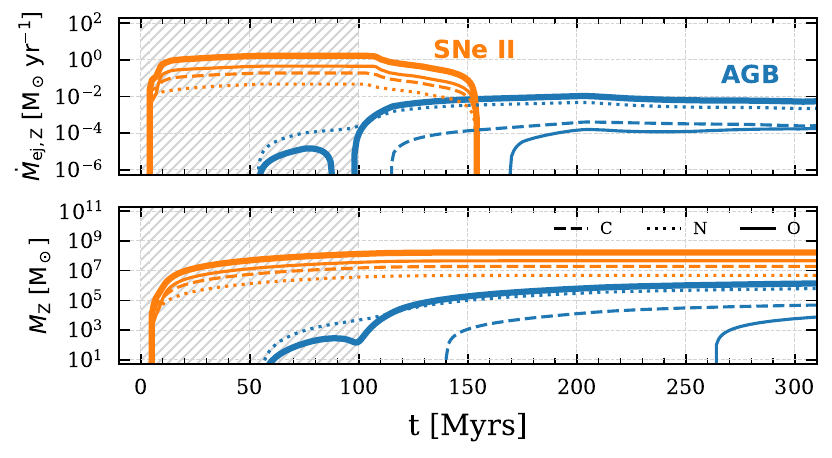}\vspace{-3mm}
\caption{Illustration of the metal build-up from our simple analytical model.
We assume a starburst forming $5\times10^9\,{\rm M_\odot}$ within $100\,{\rm Myr}$ at a rate of $50\,{\rm M_\odot\,yr^{-1}}$ (hatched area).
{\it Top:} Instantaneous metal ejection rate (${ \dot{M}_{\rm ej,Z}}$) over time from type II SNe and AGB stars for carbon (dashed), nitrogen (dotted), oxygen (solid), and their sum (thick solid).
{\it Bottom:} Cumulative ejected metal mass over time. 
\label{fig:metalmodel1}}
\end{figure}

\subsection{The Fast Build-up of Metals in the View of a Simple Analytical Model}\label{sec:discussionanalytical}

We start by implementing a simple analytical model, whose details can be reviewed in \citet{feldmann15} as well as the related papers \citep{lilly13,zahid14a,dave12}.
In brief, galaxies are treated as semi-closed boxes allowing the inflow and outflow of gas, thereby following simple continuity equations. The star formation is regulated by a star formation efficiency (or depletion time), and the metal and dust evolution is coupled to star formation by empirical yields and modified by the inflow and outflow (e.g., through feedback from stellar winds and Supernovae) of gas.
Thus the evolution of a galaxy through cosmic time is overall described by a given gas accretion rate ($\dot{M}_{\rm gas,in}$), the metallicity of the infalling gas (given by the per-cent fraction to the ISM metal-to-gas ratio; $r_{\rm Z}$), outflow rate (mass-loading factor, $\epsilon_{\rm out} \equiv \frac{\dot{M}_{\rm out}}{\rm SFR}$), and star formation efficiency (SFE) or gas depletion time ($t_{\rm depl}\propto {\rm SFE}^{-1}$).
Note that these parameters may be dependent on time, stellar mass, and star formation. For our simple demonstration purpose (and to minimize the number of variables) we treat the accretion rate and depletion time as time-independent constant factors as there are no observational indications of a strong evolution at early cosmic times \citep[e.g.,][]{scoville23}. However, the mass-loading factor $\epsilon_{\rm out}$, which controls the slope of the mass-metallicity relation, is dependent on stellar mass and redshift as discussed below.

We derive the stellar mass, SFR, gas and dust mass empirically by running the model forward in time, thereby computing these parameters in every time step

\begin{equation}\label{eq:dependencies}
    \begin{split}
M_\star(t) &= M_\star(t-\Delta t) + \dot{M}_\star(t)\,\Delta t, \\
{\rm SFR}(t) &\equiv \dot{M}_\star(t) =  \frac{M_{\rm gas}(t)}{\left<t_{\rm depl}\right>}, \\
M_{\rm gas}(t) &= M_{\rm gas}(t-\Delta t) + \left<\dot{M}_{\rm gas,in}\right>\,\Delta t,\\
M_{\rm Z}(t) &= M_{\rm Z}(t-\Delta t) + \dot{M}_{\rm Z}(t)\,\Delta t.
    \end{split}
\end{equation}

The core of this calculation is the computation of the total metal mass ($M_{\rm Z}$) via the effective metal enrichment of the ISM, which can be written as

\begin{equation} \label{eq:outflow}
\begin{split}
\dot{M}_{\rm Z}(t) &= \dot{M}_{\rm ej, Z}(t)\,-\,\zeta\,(1-R)\,\dot{M_\star}(t) \\
 & - \zeta\,\epsilon_{\rm out}(M_\star,t)\,\dot{M_\star}(t)\,+\,r_{\rm Z}\,\zeta\,\dot{M}_{\rm gas,in},
\end{split}
\end{equation}

where $\dot{M}_{\rm ej, Z}$ is the total metal ejection rate, $R=0.46$ \citep[][]{krumholz12,feldmann15}, and $\zeta~\equiv~\frac{M_{\rm Z}(t)}{M_{\rm gas}(t)}$ is the metal-to-gas mass ratio of the ISM. In other words, the metal production rate is altered by the lock-up of metals in the stars, the outflow of metal-rich gas, and the inflow of gas (which may be metal enriched or metal poor).
The oxygen abundance metallicity is then computed as $\rm 12+\log(O/H) = 8.69+\log(\zeta\,/\,0.014)$ \citep{allendeprieto01,feldmann15} using the calibration $Z_\odot \sim 0.014$ \citep{asplund09}.

For deriving the mass-loading factor (which is dependent on stellar mass and redshift), we use the results from the \textsc{EAGLE} cosmological simulation \citep{schaye15,crain15}, specifically from the work by \citet[][figure 3]{mitchell20}, which we parameterized analytically by a double power-law as described in Appendix~\ref{app:massloading}. This parameterization is consistent with the one by \citet{muratov15} based on the \textsc{FIRE} simulation \citep{hopkins14}.

While previous studies have assumed a constant yield for the production of metals, we adopt here more realistic metallicity ejection rates into the ISM.
Specifically, we used the mass-dependent metal yields computed for Asymptotic Giant Branch (AGB) stars and type II Supernovae (SNe) from \citet{dellagli19} and \citet{limongi18}, respectively, to obtain the mass ejection rates ($\dot{M}_{\rm ej,Z}$) into the ISM for carbon, nitrogen, and oxygen as a function of time for a given star formation history (SFH) and IMF\footnote{In the end, we assume to sum of these elements to contribute to metallicity.}. We include stars in the range of $1-7\,{\rm M_\odot}$ for AGB yields and $7-120\,{\rm M_\odot}$ for SNe II yields\footnote{Note that cores of stars more massive than $40\,{\rm M_\odot}$ may collapse to form black holes resulting in less metal output into the ISM. However, limiting our calculation to this mass limit has a negligible impact on the following results.} and use their mass-dependent lifetimes given in \citet{portinari98}.
Figure~\ref{fig:metalmodel1} shows the immediate mass ejection rate into the ISM (top panel) and the cumulative total ejected metal mass (bottom panel) as a function of time based on this model.
In this educational example, we assumed a simple SFH in the form of a starburst with constant SFR over $\Delta t=100\,{\rm Myr}$ (hatched area) and a total produced stellar mass of $5\times10^{9}\,{\rm M_\odot}$ (no SFR before or after burst). It is evident that type II SNe contribute the bulk of metals early on within $\sim150\,{\rm Myrs}$ after the start of the burst.
AGB stars (at lower masses) contribute most of the ejecta at $>100\,{\rm Myrs}$\footnote{Note that carbon yields in the most massive stars can be negative as the surface chemistry of these objects is mostly determined by hot bottom burning, which destroys carbon in the envelope. This causes the drop in the total metal abundance at $\sim150\,{\rm Myrs}$ after the onset of star formation.}.

This simple analytical model successfully reproduces the observed star-forming main sequence at $z\sim5$ \citep[e.g.,][]{khusanova21} and lower redshifts, assuming reasonable average values of $t_{\rm depl} = 5\times10^8\,{\rm yr}$ \citep[e.g.,][]{dessauges20} and accretion rates $\dot{M}_{\rm gas,in}=100\,{\rm M_\odot\,yr^{-1}}$ \citep[e.g.,][]{scoville23}.

\begin{figure}[t!]
\centering
\includegraphics[angle=0,width=\columnwidth]{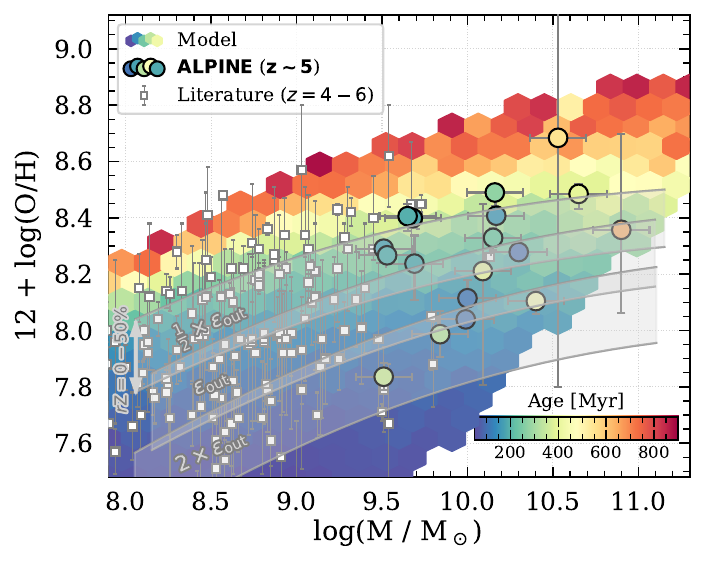}\vspace{-3mm}
\caption{
The MZR reproduced by our analytical model at $z\sim5$ (see Section~\ref{sec:discussionanalytical} for details). Model galaxies are color-coded by their age. ALPINE-CRISTAL-JWST galaxies are shown as large symbols with same color code and other literature data are shown as gray squares (see also Figure~\ref{fig:mzrobs}).
Our model captures well the range of observations and visualizes the fast metal enrichment. The upper envelope of the MZR is set by maximally old model galaxies (red).
The observed galaxy ages correlate well with the model ages.
The three gray swaths show model track ranges for two different outflow efficiencies including the nominal mass-loading factor from \citet{mitchell20}. Lower outflow efficiencies increase the metal content at a given stellar mass. The width of the swaths shows a range in the metallicity of the inflowing gas from metal free ($r_{\rm Z} = 0$) to $50\%$ ($r_{\rm Z} = 0.5$) of the metallicity of the ISM gas.
\label{fig:mzrsim}
}
\end{figure}

Figure~\ref{fig:mzrsim} shows the resulting MZR of our model at $z\sim5$ (hexagonal bins). The model output is color-coded by the age of the stellar population between $50\,{\rm Myrs}$ (blue) and $900\,{\rm Myrs}$ (red). The data from the ALPINE-CRISTAL-JWST sample are shown as large circles color-coded by age\footnote{The ages are derived from SED fitting assuming delayed $\tau$ models, see \citet{faisst20b} and \citet{tsujita26}.} on the same color scale. Other data from the literature at $z=4-6$ are shown as gray squares.
This simple analytical model with realistic metal yields is able to reproduce the parameter space of observations on the MZR at $z\sim5$ (Figure~\ref{fig:mzrsim}). In fact, our model also reproduces the FMR.
Importantly, it is also broadly consistent with most of the observed ages of the ALPINE-CRISTAL-JWST galaxies (shown at same color scale in Figure~\ref{fig:mzrsim}). We note that the measurement of ages comes with large uncertainties (factors of two or more), which could explain some of the discrepancies between the observed and model ages. Furthermore, the SED-fit ages may be younger as they capture commonly the age since the last significant starburst, while the ages output by our model are the times since first star formation.
The upper boundary of the MZR is given in our model by the maximal age of a galaxy at $z\sim5$, hence maximal cumulative metal abundance produced by our model.
The model reemphasizes the fast metal build-up; metallicities of $\rm 12+\log{(O/H)} \sim 8.3$ (the median at $3\times10^9\,{\rm M_\odot}$) are reached after only $200\,{\rm Myrs}$ of star formation. 

In addition, we can investigate the impact of gas inflows and the mass loading factor on the results. The three gray swaths in Figure~\ref{fig:mzrsim} show our analytical model with three different mass-loading implementations: the nominal $\epsilon_{\rm out}$ as well as two adjustment; $0.5\times\epsilon_{\rm out}$ and $2\times\epsilon_{\rm out}$. (Note that this is equivalent to $\epsilon_{\rm SN} = 0.5$ and $2$ in \citealt{pallottini25}.)
The width of the swaths shows a range of relative metal enrichment of the infalling gas from pristine ($0\%$) to $50\%$ of the ISM metallicity. As shown in \citet{wang26}, the CGM around the ALPINE-CRISTAL-JWST galaxies is already significantly metal enriched, therefore some recycling of the already metal-enriched gas is expected.
We find that the nominal $\epsilon_{\rm out}$ derived from the \textsc{EAGLE} (or \textsc{FIRE-2}) simulation underpredicts the metal abundance. Instead, a factor of two decreased outflow efficiency (or mass-loading factor) puts our model in better agreement with the observed MZR.
Interestingly, such a weaker outflow efficiency (lower mass-loading factor) is also more consistent with the measured mass-loading factors ($\epsilon_{\rm out} \sim 1$ at $\rm \log(M/M_\odot) \sim 10$) of the ALPINE-CRISTAL galaxies \citep{ginolfi20,birkin25} as shown in Figure~\ref{fig:epsout} in Appendix~\ref{app:massloading}.
These findings are consistent with \citet{pallottini25}, which suggests a reduction of the outflow efficiency by a factor of two-thirds ($\epsilon_{\rm SN} \sim 0.3$). 
Observationally, the effect of outflows is difficult to constrain, as it depends on various physical quantities such as star formation, outflow velocity, or density of the ISM, which may all depend on redshift \citep[e.g.,][]{davies19}.

In summary, our simple analytical model describes well the overall MZR at $z\sim5$ of low-mass JWST-detected and massive ALPINE-CRISTAL-JWST galaxies.
The range of observed metallicities at a given stellar mass (hence the MZR scatter) can, to first order, be explained by a mix of different stellar population ages, outflow efficiencies (mass-loading factors), and the metal enrichment of infalling gas. Notably, metals build up quickly to the observed levels within $<200\,{\rm Myrs}$ since the onset of star formation.
To reproduce the most metal-poor systems, either very young ($\ll200\,{\rm Myrs}$) stellar populations or efficient metal-rich mass outflows (Figure~\ref{fig:mzrsim}) need to be assumed.
Specifically, due to the fast build-up of metals, any previous metals need to be removed in order to observe a galaxy as metal-poor. This can be achieved by strong outflows or the accretion of pristine gas. In addition, metal poor systems could be associated with under-dense environments (isolating the galaxies from growth through gas inflows), such as suspected for a recently found metal-poor system {\em AMORE6} at $z=5.7$ \citep{morishita25}.
The most metal-rich systems are either old or have significantly less efficient outflows.
Future, more detailed, observations of outflow rates from optical emission lines and the large-scale structure of the surrounding environment of these galaxies may help to distinguish these processes.
In Section~\ref{sec:fmrscatter} we will further study the effect of a stochastic SFH, which may cause galaxies to scatter to lower metallicity.

Finally, we note that the observed metallicities are spatially integrated (consistent with the output of our analytical model). Spatially resolved metal abundances may show additional large variations. For example, pockets of recent star formation driven by pristine gas inflow may have lower-than-average metallicities \citep{sanchezalmeida14,sanchezalmeida15}.
This is also further discussed in \citet{fujimoto25} and \citet{lopez25}.

\input{table3.tex}

\subsection{Scatter of the FMR and its Universality}\label{sec:fmrscatter}

As shown in the right panel of Figure~\ref{fig:fmr}, there is significant scatter in the FMR especially at lower stellar masses. For a $\mu_0=0.66$ \citep{andrews13}, we measure a scatter of $\sigma_{\rm FMR} \sim 0.24\,{\rm dex}$ based on the data at $z=4-6$. Similar to \citet{mannucci10}, we have additionally tried to minimize the FMR scatter by choosing an optimal $\mu_0$. We found that $\mu_0=0.22$ decreases the scatter to $\sigma_{\rm FMR} \sim 0.22\,{\rm dex}$, which is negligibly smaller than for $\mu_0=0.66$. The value of the scatter is consistent with \citet{curti24} at similar redshifts. We therefore use $\mu_0=0.66$ in the following to be consistent with other studies \citep{andrews13,curti20}.
The intrinsic FMR scatter is $\hat{\sigma}_{\rm FMR} = \sqrt{\sigma_{\rm FMR}^2 - \sigma_{\rm meas}^2}\sim 0.10-0.13\,{\rm dex}$, for $\mu_0=0.22-0.66$ and assuming a conservative uncertainty in the metallicity measurement of $\sigma_{\rm meas}=0.2\,{\rm dex}$ \citep{curti24}.
This is a factor of five larger than the scatter measured at $z=0$, $\hat{\sigma}_{\rm FMR,0} \lesssim \sigma_{\rm FMR,0} = 0.02$ \citep{mannucci10}, and also significantly larger than at $z\sim2-3$, $\hat{\sigma}_{\rm FMR} \sim 0.06$ \citep{sanders21}. Note that the latter finds that $\mu_0 = 0.60$ minimizes the scatter for sample at $z=2.3$ and $z=3.3$. Measurements of the FMR scatter at different redshifts are summarized in Table~\ref{tab:fmrscatter}. 
As mentioned previously, the large scatter in the FMR at $z=5$ cannot be minimized by adjusting $\mu_0$. Furthermore, the scatter is similar to the scatter of the MZR (corresponding to $\mu_0 = 0$). This questions the existence of a {\em fundamental} metallicity relation at $z=5$, which has also been discussed in previous works \citep[e.g.,][]{steidel14,korhonen-cuestas25}. The ``non-reducible'' scatter of the MZR when including star formation as a third parameter therefore hints towards a difference compared to lower redshifts samples --- for example a bursty star formation or early metal build up --- which is explored in the next section.

\subsection{Can a Bursty Star Formation Explain the Scatter of the FMR?}\label{sec:bursty}

It is now well established that star formation is stochastic and bursty especially for galaxies at high redshifts and low stellar masses \citep[e.g.,][]{sun25,faisst24,endsley24,navarrocarrera24,ciesla24,munoz23,shenXuejian23,asada23,mehta23,faisst19,emami19,caplar19,weisz12}.
This stochasticity may be driven by the outflow of gas through SNe feedback (specifically at low stellar masses), changes in gas accretion, and gas compaction and depletion events \citep[e.g.,][]{fauchergiguere18}.

We further tested this scenario by directly using a stochastic SFH in our analytical model described in Section~\ref{sec:discussionanalytical}. Note that at this point, the model does not assume an equilibrium state anymore (contrary to the common ``gas regulator'' or ``bathtub'' model), but we directly compute the metal enrichment for an arbitrarily changing SFH.
We adopt a stochastic (bursty) SFH as described in \citet{faisst24} and based on the mathematical prescription by \citet{caplar19}. Specifically, we use the \texttt{DELightcurveSimulation} Python package\footnote{\url{https://github.com/samconnolly/DELightcurveSimulation}} \citep{connolly16}, which is based on the \citet{emmanoulopoulos13} light-curve simulation algorithm. We assume a generalized power spectrum density \citep[PSD, with a slope $\alpha_{\rm PSD}=2$;][]{faisst24}. The stochastic SFH is parameterized by a burstiness amplitude ($\sigma_{\rm MS}$; {\it i.e.} the amplitude distance from the main-sequence in dex) and a correlation time ($\tau_{\rm break}$; {\it i.e.} roughly the $\Delta t$ between amplitude peaks).
We simulated SFHs with $\sigma_{\rm MS}=0.1$, $0.3$, and $0.6\,{\rm dex}$ and $\tau_{\rm break}=5$, $50$, and $200\,{\rm Myrs}$. The latter is expected from simulations such as SPHINX$^{20}$ \citep{rosdahl18,rosdahl22,katz22,faisst24}.

\begin{figure}[t!]
\centering
\includegraphics[angle=0,width=1\columnwidth]{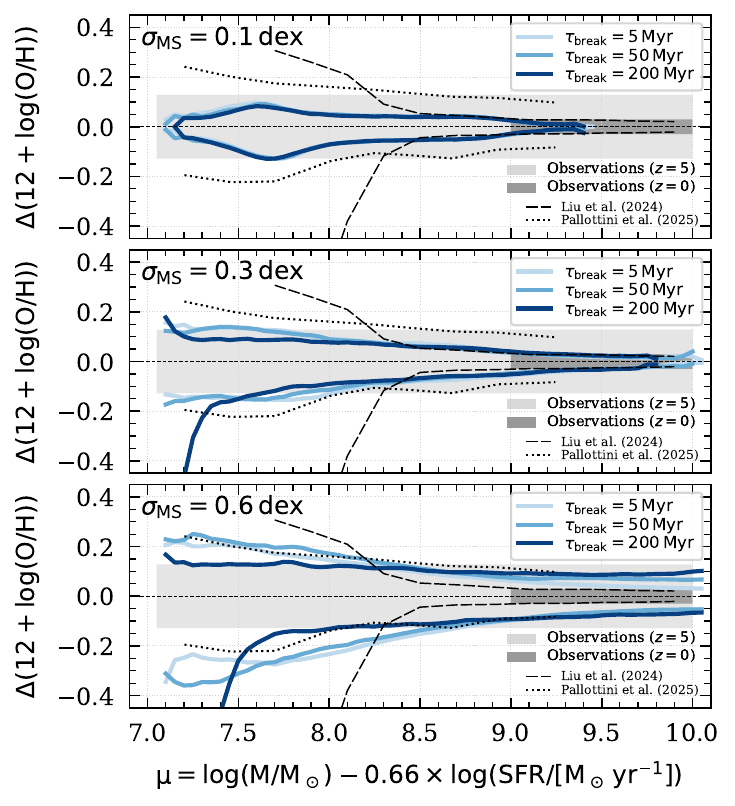}
\vspace{-3mm}
\caption{Scatter of the FMR as a function of $\mu$ for different assumption of burstiness of the SFH. The SFH is described by a burstiness amplitude ($\sigma_{\rm MS}$, different panels) and a correlation timescale ($\tau_{\rm break}$, shades of blue). The observed FMR scatter (deconvolved by measurement errors) at $z\sim5$ and $z=0$ is shown in light and dark gray, respectively. We also show the scatter derived from the analytical models from \citet{liulunjun24} (long-dashed line) and \citet{pallottini25} (dotted line).
\label{fig:fmrscatter}}
\end{figure}

The resulting scatter of the FMR from this model as well as other analytical models (see Section~\ref{sec:discussioncosmological}) is compared to the observed scatter ($\hat{\sigma}_{\rm FMR}$) in Figure~\ref{fig:fmrscatter}.
An increased burstiness amplitude increases the FMR scatter, but we do not find a significant dependency on $\tau_{\rm break}$ except for the largest $\sigma_{\rm MS}$, where a shorter correlation time causes larger scatter.
The scatter is largest at low stellar masses (corresponding to early times in a galaxy's evolution and low $\mu$) and decreases towards higher stellar masses.
Furthermore, increased burstiness causes a {\it negative} (downward) scatter on the FMR similar to observational findings.
The increased (downward) scatter could be explained as follows: as the galaxies are young and pristine, small variations in metal abundances (caused by star formation burstiness and almost instantaneous SNe dust production) can be observed causing the observed scatter. Further infall of pristine gas may lower the gas-phase metal abundance temporarily. As the galaxies evolve, they become more metal enriched, producing metals more steadily (e.g., through older AGB stars). Local fluctuations in metal enrichment due to SNe metal production will be below the sensitivity limit of current observations, thus the overall observed scatter is reduced.
We note that in our simple model the burstiness is constant with stellar mass \citep[in reality, more massive galaxies are expected to be less bursty, e.g.][]{weisz12,emami19}. The reduction of the observed scatter can therefore be explained purely due to the metal build-up (decreasing the contrast with respect to small metal abundance fluctuations) and the importance of more smooth metal production mechanisms (AGB metal production vs. SNe metal production).
We find that a burstiness amplitude of $\sigma_{\rm MS} = 0.6\,{\rm dex}$ can reproduce the observed scatter (deconvolved by measurement errors) well at high redshifts. However, a lower amplitude ($\ll0.3\,{\rm dex}$) would be sufficient to explain the FMR scatter of low redshift galaxies.
Our findings are consistent with a study by \citet{pallottini25}, which shows, using a similar model, that the scatter of the MZR significantly depends on the burstiness of star formation.
Furthermore, studies at low redshifts suggest extremely metal poor galaxies to be dispersion dominated and gas rich, suggesting they are dominated by bursts and gas inflow \citep[e.g.,][]{isobe23}.

We conclude that the scatter of the FMR at low metallicities (hence specifically low stellar masses) may be due to a bursty star formation history. Similar to observations, our bursty model suggests an asymmetric and negative FMR scatter. At later times, however, burstiness does not have a significant impact on the total observed metallicity. This may explain the smaller FMR scatter at low redshifts compared to higher redshifts. Furthermore, metal mixing may play a role.

\subsection{Comparison to Simulations and other Analytical Models}\label{sec:discussioncosmological}

Our simple analytical model described in the previous sections reproduces well the observations at $z\sim5$ of the FMR and MZR. Furthermore, we showed that the {\it asymmetric} scatter in the FMR can be naturally explained by a stochastic star formation history.
However, our simple model only provides limited predictions. For example, the burstiness of star formation is connected to (or originates from) other physical processes such as gas outflows and star formation efficiencies or change in the gas inflows.
In this section, we therefore compare the observations to predictions from cosmological simulations that include realistic feedback, outflow, inflow, and mass assembly.

\begin{figure*}[t!]
\centering
\includegraphics[angle=0,width=1\textwidth]{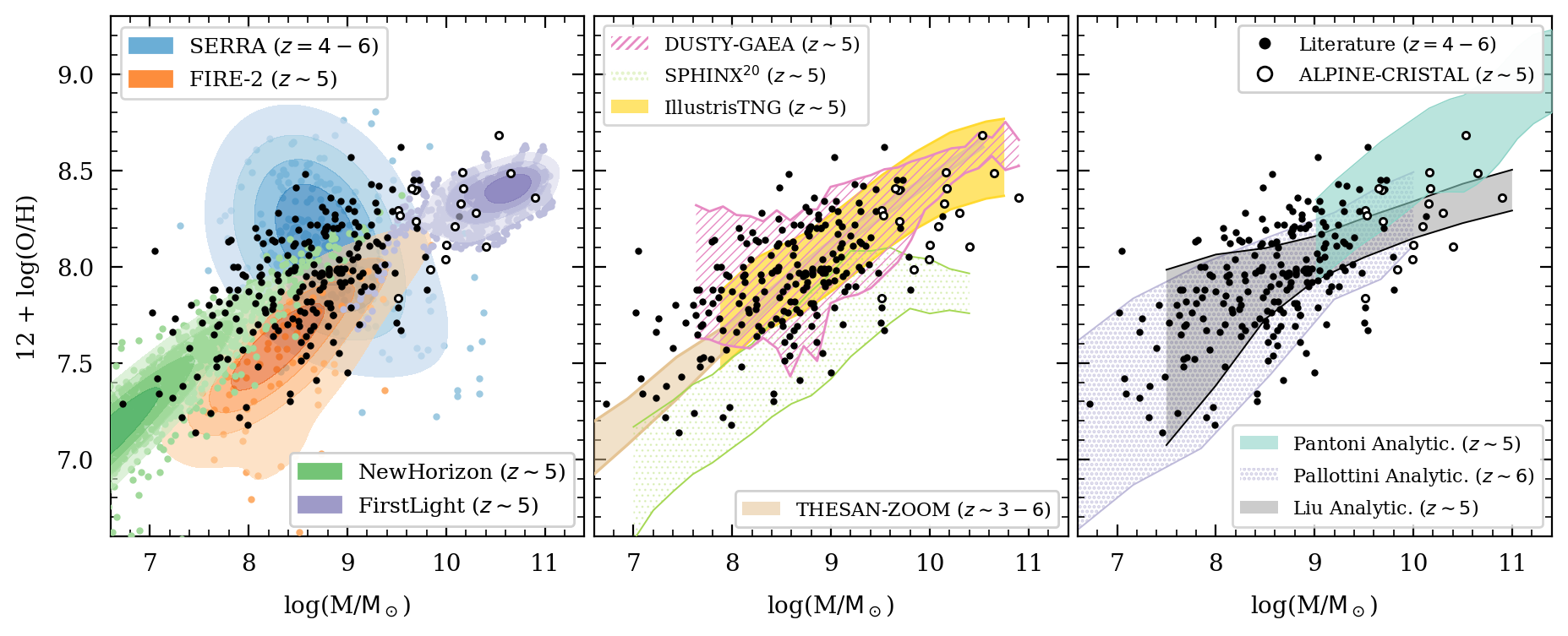}\vspace{-3mm}
\caption{
Comparison of the observed MZR at $z\sim5$ (black circles, see also Figure~\ref{fig:mzrobs}) with various simulations. For better visibility, we show three panels focusing on different simulations.
{\em Left:} The different contours show results from SERRA \citep[blue;][]{pallottini22}, \textsc{FIRE-2} \citep[orange;][]{ma16,marszewski24}, \textsc{NewHorizon} \citep[green;][]{dubois21}, and \textsc{FirstLight} \citep[purple;][]{Ceverino2017,nakazato23,ceverino26}.
{\em Middle:} The different areas show results from \textsc{DUSTY-GAEA} \citep[pink forward line-hatched;][]{osman25}, \textsc{SPHINX}$^{20}$ \citep[green dotted;][]{rosdahl18,rosdahl22,katz23}, \textsc{IllustrisTNG} \citep[$z=4-6$; gold shaded;][]{hirschmann23}, and \textsc{THESAN-ZOOM} \citep[salmon shaded;][]{kannan25}.
{\em Right:} Results from different analytical models are shown including \citet{pantoni19} (teal shaded), \citet{pallottini25} (purple dotted), and \citet{liulunjun24} (gray shaded).
\label{fig:mzrcossim}
}
\end{figure*}

Figure~\ref{fig:mzrcossim} shows the observations at $z\sim4-6$ (black dots and circles) compared to various cosmological simulations and analytical models which are listed in the following.

\paragraph{\textsc{SERRA} Zoom-In Simulation~}~ The \textsc{SERRA} simulations are a set of zoom-in high-resolution ($1.2\times 10^4\,{\rm M_\odot}$, $\simeq 25\,{\rm pc}$ at $z=7.7$) cosmological simulations including non-equilibrium chemistry and on-the-fly radiative transfer \citep{pallottini22}. The masses of the galaxies range approximately between $10^7 - 5\times10^9\,{\rm M_\odot}$. Generally, the \textsc{SERRA} galaxies reproduce well the observed population in the mass range $10^8-10^9\,{\rm M_\odot}$, but they slightly overpredict and underpredict the metal abundances at lower and higher stellar masses, respectively (left panel, blue contours). The underprediction of metals in massive galaxies may be tied to the stochasticity of the \textsc{SERRA} galaxies as discussed in \citet{pallottini25}. The overprediction at lower masses could be due to weak feedback \citep[c.f. ][]{lunjunliu26}.

\paragraph{\textsc{FIRE-2} Zoom-In Simulation~}~\textsc{FIRE-2} is a suite of cosmological zoom-in simulations including the implementation of realistic stellar and radiation feedback and winds \citep{hopkins14,hopkins18,ma18a,ma18b,ma19,hopkins23}. The particle resolution is $100-7000\,{\rm M_\odot}$ and the masses of the galaxies range from $10^6-10^{10}\,{\rm M_\odot}$ at $z\sim5$. The MZR derived in \citet{marszewski24} \citep[see also][]{ma16} compares well to observations, although it slightly underpredicts the metal abundances at $<10^8\,{\rm M_\odot}$ (left panel, orange contours).

\paragraph{\textsc{NewHorizon} Zoom-in Simulation~}~The \textsc{NewHorizon} project \citep{dubois21} performs a high-resolution zoom-in simulation extracted from the parent simulation \textsc{Horizon-AGN} \citep[$(142\,{\rm Mpc})^3$;][]{dubois14} in a box size of $(16\,{\rm Mpc})^3$ and a resolution of up to $34\,{\rm pc}$. These simulations are generally in good agreement with the observed MZR at stellar masses of $<10^9\,{\rm M_\odot}$ (left panel, green contours). However, the scatter is narrow and the sample size at higher stellar masses is small, both likely due to the small volume probed by the simulation.

\paragraph{\textsc{FirstLight} Zoom-in Simulation~}~\textsc{FirstLight} is a suite of multi-object zoom-in simulations including in total 430 galaxies \citep{Ceverino2017}. The simulations are assorted in four different volumes from $10$ to $80\,{\rm Mpc/h}$ with different mass resolutions. To match the massive $M_*\sim10^{10-11}\,{\rm M_\odot}$ ALPINE-CRISTAL galaxies at $z\sim5$, we make use of the two largest box sizes \citep[$40$ and $80\,{\rm cMpc/h}$;][]{nakazato23,ceverino26} with dark matter mass resolutions of $8\times10^4$ and $6\times10^8\,M_\odot$, respectively, and maximal spatial resolutions of $17$ and $35\,{\rm pc}$, respectively. These simulations are in good agreement with the ALPINE-CRISTAL-JWST observations (left panel, purple contours).

\paragraph{\textsc{DUSTY-GAEA} Simulation~}~\textsc{DUSTY-GAEA} is a modified version of the state-of-the-art semi-analytic GAlaxy Evolution and Assembly \citep[GAEA;][]{delucia14b,hirschmann16,xie20,fontanot20,delucia24} model, where a dust physical prescription is included in a self-consistent manner \citep{osman25}. The \textsc{DUSTY-GAEA} model is run on the \textsc{Millennium} \citep{springel05} and \textsc{Millennium II} \citep{boylankolchin09} merger trees to cover the wide range of the stellar mass ($\rm \log(M/M_\odot) > 7$) on the MZR plane. This model describes well the observed MZR (middle panel, pink forward-hatched).

\paragraph{\textsc{SPHINX$^{20}$} Simulation~}~The \textsc{SPHINX$^{20}$} is a full box cosmological radiation hydrodynamics simulation optimized to simulate cosmic reionization and the multi-phase ISM of high-$z$ galaxies to provide statistical samples of galaxies similar to those currently being observed by JWST \citep{rosdahl18,rosdahl22,katz23}. The metallicity measurement are taken from the current data release described in \citet{katz23}. These simulations generally underpredict the metal abundances of $z\sim5$ galaxies (middle panel, green dotted) and occupy a similar parameter space as the \textsc{FIRE-2} galaxies in terms of stellar mass and metallicity.

\paragraph{\textsc{IllustrisTNG} Simulation~}~\textsc{IllustrisTNG} is a suite of large volume, cosmological, gravo-magnetohydrodynamical simulations, covering $302.6^3$, $106.5^3$, and $51.7^3\,{\rm cMpc^3}$, respectively, and particle resolutions between $10^4-10^7\,{\rm M_\odot}$ \citep{marinacci18,naiman18,nelson18,nelson19,springel18,pillepich18,pillepich19}. We compare the observations to the resulting MZR for stellar masses of $10^8-10^{11}\,{\rm M_\odot}$ from \citet{hirschmann23}, who derived accurate optical emission line for the simulated galaxies in post-processing (middle panel, yellow shaded). The predicted metal abundances reproduce well the observations.

\paragraph{\textsc{THESAN-ZOOM} Zoom-In Simulation~}~This is a zoom-in to the cosmological simulation \textsc{THESAN-1} \citep{kannan22,smith22,garaldi22} at a baryonic mass and spatial resolution of $142\,{\rm M_\odot}$ and $\sim17\,{\rm cpc}$, respectively. It includes self-consistently the effects of supernova feedback, radiation fields, dust physics, and low temperature cooling via molecular hydrogen \citep{marinacci19,kannan20,kannan21}. We use the MZR prediction presented in \citet{kannan25} and find a good agreement with the observed relation (middle panel, salmon shaded). However, we notice that the spread in metallicity is significantly narrower than that observed.

\paragraph{\citet{pantoni19} Analytical Model~}~Similar to our model, this analytical model self-consistently describes the spatially averaged time evolution of the gas, stellar, metal, and dust contents \citep[see also][]{lapi20}, based on the model by \citet{lapi17}. It is thought to better represent high-$z$ galaxies and progenitors of early type galaxies compared to constant gas inflow models. It reproduces well the overall metal abundances, however, it slightly overestimates the metallicity of the most massive galaxies (right panel, teal shaded).

\paragraph{\citet{pallottini25} Analytical Model~}~ This minimal physical model, directly tied to the dark matter halo growth, is similar in general to the one used in this work. The model has adjustable SNe delay, feedback, and variable accretion rate to simulate an oscillating, bursty SFH. Here we use the model ``weak feedback and modulated accretion'', which reproduces the observed MZR well (right panel, purple dotted). As shown in Figure~\ref{fig:fmrscatter}, this model leads to a similar scatter in the FMR as our stochastic SFH model assuming $\sigma_{\rm MS}$ scatter between $0.3-0.6\,{\rm dex}$ and recovers well the observed FMR scatter at $z\sim5$ (however, it slightly overpredicts the FMR scatter at the lowest stellar masses). Note that the scatter reported here from \citet{pallottini25} is the maximal range of models (not the $1\sigma$ value). Importantly, a decreasing trend of FMR scatter with stellar mass is reported as well by this model.

\begin{figure*}[t!]
\centering
\includegraphics[angle=0,width=0.48\textwidth]{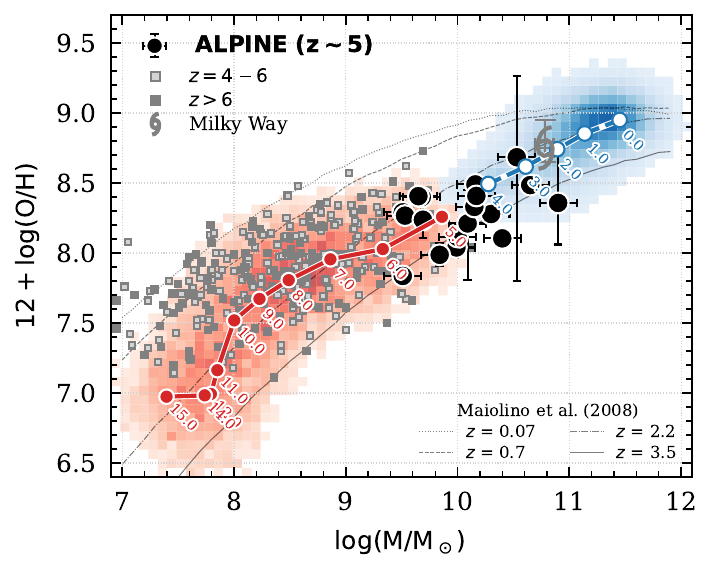}
\includegraphics[angle=0,width=0.48\textwidth]{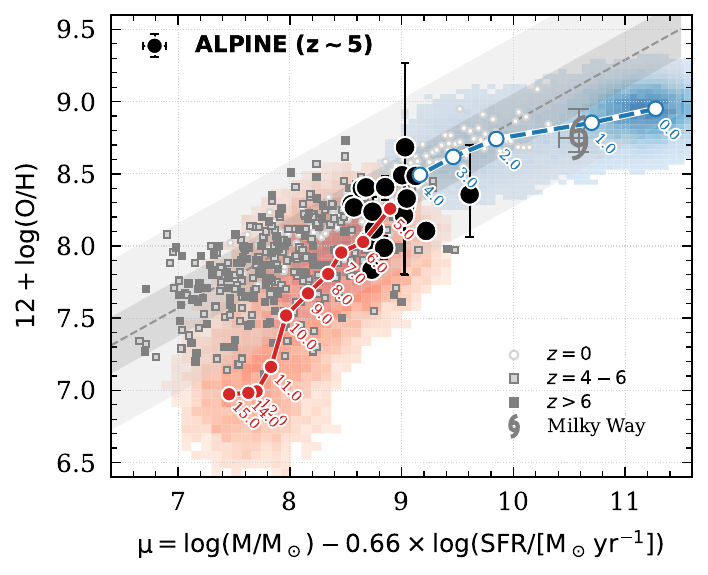}
\vspace{-3mm}
\caption{Tracks of simulated DUSTY-GAEA model progenitors (red) and descendants (blue) of the ALPINE-CRISTAL-JWST sample.
{\it Left:} The MZR with observations at $z=4-6$ (gray squares) and $z>6$ (solid dark gray squares) as well as our Milky Way ($\S$) indicated. The ALPINE-CRISTAL-JWST sample is indicated with large black circles. The MZR parameterizations from \citet{maiolino08} at different redshifts are indicated as thin lines. The DUSTY-GAEA models (matched to the ALPINE-CRISTAL-JWST galaxies in metallicity, SFR, and stellar mass) are shown in red ranging from $z=5$ to $z=15$ ($\Delta t = 1\,{\rm Gyr}$, redshifts indicated by red dots) and blue from $z=5$ to $z=0$. The red/blue cloud shows the possible range of progenitor/descendant tracks.
{\it Right:} Same as left panel, but showing the FMR. The local FMR \citep{andrews13} is shown as dashed line. The $1\sigma$ and $3\sigma$ scatter of the $z=4-6$ observations around this local relation are shown as light and dark gray bands, respectively. 
\label{fig:gaeasim}}
\end{figure*}

\paragraph{\citet{liulunjun24} Analytical Model~}~This physically motivated analytical model takes a similar approach as \citet{pallottini25}. It implements a modulated SFH as a result of delayed injection of supernova feedback \citep{furlanetto22} in a self-regulated disk model as the basis for describing galaxy-scale star formation \citep[see also][]{fauchergiguere13,fauchergiguere18}. It self-consistently models the gas, stellar, metal contents, and line emissions that trace the star formation. Adopting the scenario with galactic wind driven by reduced supernova feedback (smaller mass-loading factor), this model matches the measured MZR (right panel, gray shaded) and is broadly consistent with the observed $1\sigma$ scatter of the FMR derived from our stochastic SFH model (Figure~\ref{fig:fmrscatter}).
However, the scatter of metallicity predicted by this model is larger than observed at stellar mass of $<10^8 M_{\odot}$. This can be induced by the burstiness in low-mass galaxies whose shallow gravitational potentials are unlikely to prevent metal contents from being evacuated by feedback-driven outflows.

\vspace{3mm}
We note that most simulations reproduce the scatter of the observe metal abundances at a given stellar mass. This shows that the scatter in the MZR is dominated by physical processes and not measurement uncertainties (see also Section~\ref{sec:discussionanalytical}).
The reason that some cosmological simulations underpredict the metal abundances could be due to limited volume coverage to simulate galaxies that have above-average metal enrichment (e.g., fast evolving galaxies). For example, the \textsc{FIRE-2} simulation covers $\sim(43\,{\rm cMpc})^3$ and the SPHINX$^{20}$ simulation covers $\sim(20\,{\rm cMpc})^3$, which are both significantly smaller volumes than covered by observations ($\sim(100\,{\rm cMpc})^3$). This may be connected to failing to reproduce the significant ALMA far-infrared \oiii$_{88\,{\rm \mu m}}$ emission in two $z>12$ galaxies \citep[see discussion in ][]{yang25}.
However, other differences such as the implementation of feedback, radiative heating and cooling, or chemical models may also change the resulting metal enrichment.
Lastly, observational constraints may still hinder the detection of low-metallicity galaxies at high redshifts.
This could be mitigated by observing lensed high-redshift galaxies \citep[e.g.,][]{welch25,christensen12a,christensen12b,vanzella23}.

\subsection{The progenitors and descendants of the ALPINE galaxies}

We use the \textsc{DUSTY-GAEA} simulation model \citep{osman25} to characterize the evolutionary path of the ALPINE-CRISTAL-JWST galaxies on the MZR and FMR plane.
We chose \textsc{DUSTY-GAEA} because it is able to reproduce well the metal abundances over the full stellar mass range of galaxies at $z\sim5$ (Figure~\ref{fig:mzrcossim}) as well as lower redshifts \citep{fontanot24}.

We chose $\sim2400$ representative galaxies in the DUSTY-GAEA model that cover the range of metallicity ($\rm 12+\log(O/H) = 7.8 - 8.7$), SFR ($\rm \log(SFR/[M_\odot\,yr^{-1}])=1.2-2.3$), and stellar mass ($10^{9.5}-10^{10.9}\,{\rm M_\odot}$) of the ALPINE-CRISTAL-JWST sample. We then trace the galaxies backward and forward in time by following their merger trees. We use both Millennium \citep{springel05} and Millennium II \citep{boylankolchin09} merger trees to cover the wide range of the stellar mass ($\rm \log(M/M_\odot) > 7$).

The left panel of Figure~\ref{fig:gaeasim} shows the MZR including observations at $z=4-6$ (ALPINE-CRISTAL-JWST galaxies indicated with large black symbols). The simulated \textsc{DUSTY-GAEA} tracks of the ALPINE-CRISTAL-JWST progenitor galaxies are indicated as red lines from redshifts $z=15$ to $z=5$ (indicated by red dots). The blue line shows the tracks of the descendants from $z=5$ to $z=0$ (indicated by blue dots). The red/blue cloud shows the range of galaxy model progenitors/descendants marginalized over all those simulated galaxies. The tracks themselves connect the median values at a given redshift snapshot.
The progenitors of the ALPINE-CRISTAL-JWST galaxies follow a slightly sloped MZR to $z\sim10$ ($\sim700\,{\rm Myrs}$ in the past) after which there are indications that the relation drops off more steeply to lower metallicities (note that the time difference from $z=10$ to $z=15$ is only $\sim250\,{\rm Myrs}$).

\begin{figure*}[t!]
\centering
\includegraphics[angle=0,width=1\textwidth]{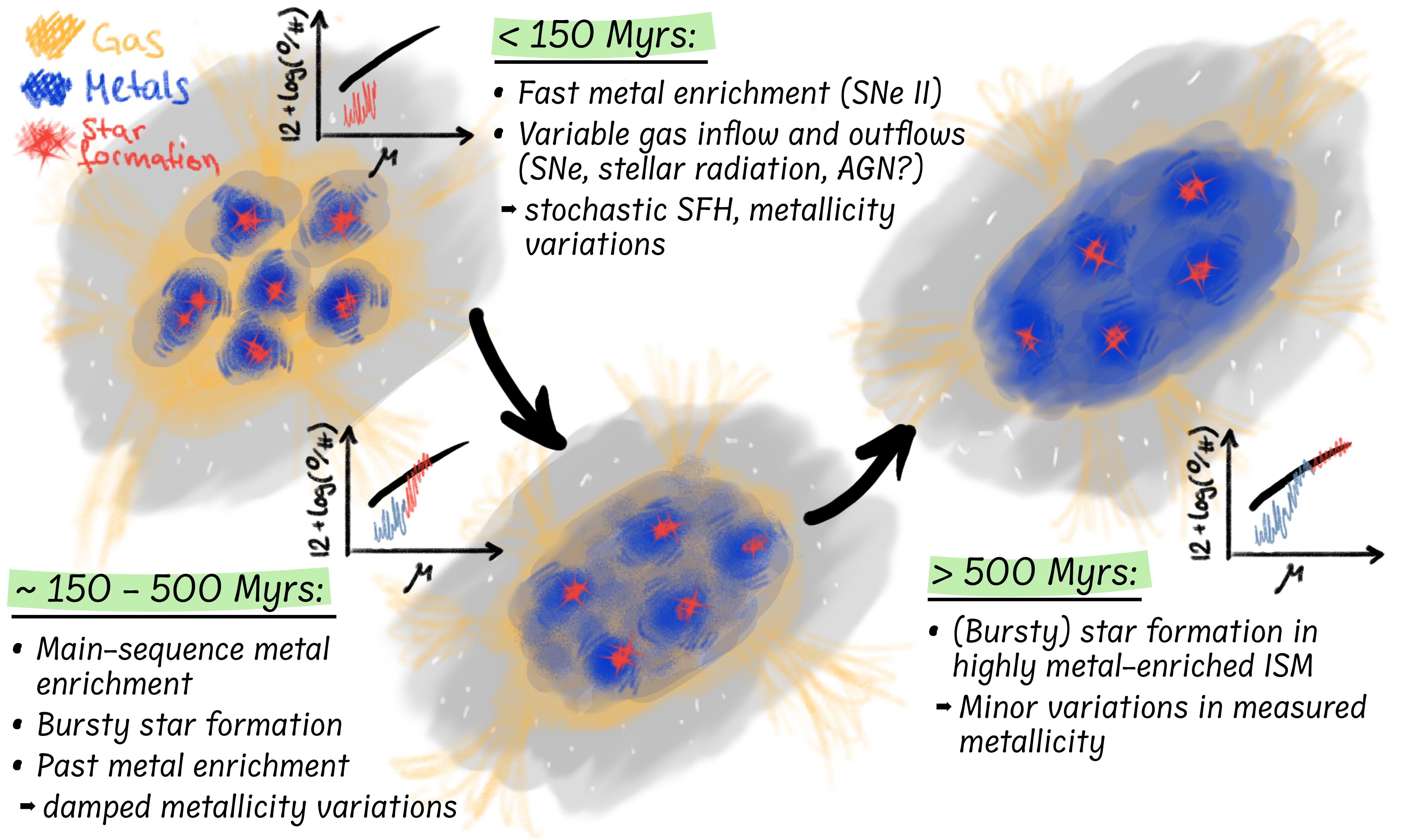}\vspace{-3mm}
\caption{
Sketch of the early metal enrichment of galaxies. Gas (yellow), metals (blue), and areas of active star formation (red) are indicated. Metal enrichment is indicated as blue arcs and gas inflow as yellow streams. The inset figures show the current location and scatter (red) of the galaxy at a given time on the FMR.
Young galaxies experience fast metal enrichment with large scatter on MZR and FMR due to stochastic star formation caused by variable gas inflows and outflows. Older galaxies continue to have bursty star formation but with damped MZR/FMR scatter due to past metal enrichment. 
\label{fig:summaryfigure}
}
\end{figure*}

One caveat is that the \textsc{Millennium} simulation only sparsely populates the regime at $z>9$. We therefore compared the progenitor tracks to $\sim400$ galaxies from the Planck Millennium simulation \citep[\texttt{PMILL};][]{baugh19}. This $N$-body simulation is updated with the recent Planck cosmology \citep{planck14} and supports a higher resolution and better time sampling at high redshifts than the \textsc{Millennium} simulation. Making use of this increased sampling we found a slightly less steep metallicity evolution beyond $z=9$, but overall consistent with the previous results.

The comparison to simulations emphasizes the fast build-up of metals at early times before flattening off to join a more mature main-sequence evolution of metals over time, characterized by the local FMR. Note that this is consistent with our simple analytical model (Figure~\ref{fig:mzrcossim}), showing younger galaxies at lower metallicity at a given stellar mass. Furthermore, such a transition has been discussed in \citet{burgarella25}, suggesting a possible shift in the mechanism of metal and dust production around a critical metallicity of $12+\log{\rm (O/H)}\sim7.6$ from being dominated by stellar-dust production (mainly from SNe) to grain growth through gas-dust accretion
in the ISM.

This early build-up of metals is further visualized on the FMR (right panel, Figure~\ref{fig:gaeasim}). The progenitors of the ALPINE-CRISTAL-JWST galaxies start to ``fall off'' the local relation around $z\sim10$ as they have a significantly ($>3\sigma$, indicated by the gray bands) lower metallicity at a given $\mu$ compared to the $z=4-6$ observed population and the local relation. This suggests that the ALPINE-CRISTAL-JWST progenitors join the fundamental metallicity relation to proceed, after a rapid build-up of metals, with more mature metal build-up. Again, we note that for the \textsc{Millennium} simulation, only $22$ snapshots were saved at redshift $>5$. Therefore our statements about the rapid growth of metals at early times should be interpreted with caution. Nevertheless, from preliminary work based on simulations with a much larger number of snapshots (such as \texttt{PMILL}), results are qualitatively similar to those shown and discussed here. We will address this more explicitly in a forthcoming work \citep{osman25}.

Based on the \textsc{DUSTY-GAEA} models, the $z\rightarrow0$ descendants of the ALPINE-CRISTAL-JWST sample galaxies are massive ($\rm \log(M_\star/M_\odot)\sim 11.4$) with (super-solar) metallicities of $\rm 12+\log(O/H)\sim 9$.
The descendants follow well the expected FMR until $z\sim2$. After that, they become quiescent and their SFR drops from $\sim 60\,{\rm M_\odot\,yr^{-1}}$ ($z=2$) to $\sim 2\,{\rm M_\odot\,yr^{-1}}$ at $z=0$. At that point they leave the FMR of star-forming galaxies as shown on the right panel of Figure~\ref{fig:gaeasim}.

As comparison, we also show the approximate location of our own Milky Way in Figure~\ref{fig:gaeasim}. The total stellar mass ($6.08\pm1.14\times10^{10}\,{\rm M_\odot}$) is taken from \citet{licquia15} and the SFR ($2.0\pm0.7\,{\rm M_\odot\,yr^{-1}}$) from \citet{elia22}. The Milky Way's gas-metallicity is less constrained due to different calibrations and radial dependence. Here we assumed $\rm 12+\log(O/H) \sim 8.75$ \citep{daflon04,pilyugin06,molla19}.
The descendants of the ALPINE-CRISTAL-JWST galaxies are predicted to be five times more massive and have $\sim0.2\,{\rm dex}$ higher oxygen abundances.
However, we note most of the metals at $z<1$ are bound in stars, which would mean that the actual ISM metallicities of the simulated ALPINE $z=0$ galaxies are lower \citep{peroux20}.

In conclusion, the simulations show a consistent picture in which the ALPINE-CRISTAL-JWST galaxies build up their metals quickly in the first $\sim200\,{\rm Myrs}$ before joining the local fundamental relation around $z=10$.
Figure~\ref{fig:summaryfigure} summarizes the results from this paper in a schematics.
We now observe the mature evolution of the metal abundances, moving towards solar metallicity at the highest masses at $z=5$. This evolution continues down to $z=0$.
In other words, to witness the early build-up of metals, galaxies at $z>10$ need to be observed, or alternatively much younger galaxies at later times.
The first attempts to find these very metal-poor systems at high redshifts are already underway \citep[e.g.,][]{vanzella23,nishigaki23,cullen25,willott25,nakajima25,hsiao25,morishita25,sui26}, complementing findings of extremely metal-poor galaxies at low redshifts \citep[e.g.,][]{shi14,senchyna19,izotov18,izotov19,kojima20,pustilnik20,pustilnik21,laseter22,isobe23,karachentsev23,breneman25}.

\section{Summary \& Conclusions}\label{sec:conclusions}

We have used the ALPINE-CRISTAL-JWST survey data to investigate the spatially integrated metal enrichment of $18$ massive ($\rm \log(M_\star/M_\odot) = 9.5 - 11$) galaxies at $z\sim5$. This sample is complementary to previous studies of JWST-observed samples that probe lower stellar masses, hence allows insights into the metal enrichment of the massive early galaxies.
The metallicities are derived using the strong-line method \citep{faisst26} applying the \citet{sanders24} calibration. We find good agreement with $T_e$-based metallicity estimates for five galaxies for which such measurements are possible. The $T_e$ values vary at $\sim 1-2\times10^4\,{\rm K}$ and electron densities are between $n_e =200-1000\,{\rm cm^{-3}}$ with an average of $400\,{\rm cm^{-3}}$.
The main observational findings of our study are the following:

\begin{itemize}
    \item Corroborating on the full stellar mass range of samples from the literature and from the ALPINE-CRISTAL-JWST sample, we found a relatively shallow mass-metallicity relation at $z\sim5$ (Equation~\ref{eq:mzr}). This suggests a significant (and fast) metal enrichment even at low stellar masses.
    Overall, massive $z=5$ galaxies increase their metal abundance from $\sim40\%$ to $\sim60\%$ to cosmic noon at fixed stellar mass (Figure~\ref{fig:mzrobs}).

    \item Galaxies at $z\sim5$ are to first order consistent with the local FMR. To second order, the $z\sim5$ FMR shows a downward scatter in metallicity at a fixed stellar mass (or $\mu$) compared to the local relation (Figure~\ref{fig:fmr}). This could indicate the early phases of metal enrichment.

    \item We find the intrinsic scatter of the FMR at $z\sim5$ to be approximately five times larger compared to the local FMR (Figure~\ref{fig:fmrscatter} and Table~\ref{tab:fmrscatter}). 

\end{itemize}

We investigated these findings further in the light of a simple analytical model. Our model is similar to previous ``gas-regulator'' prescriptions \citep[e.g.][]{lilly13,feldmann15}, however, instead of using a constant yield, we have implemented a metal enrichment based on more realistic metal yields of type II SNe and AGB stars (Figure~\ref{fig:metalmodel1}). Furthermore, we use stellar mass and redshift-dependent mass-loading factors derived from the \textsc{EAGLE} simulation. 
Comparing the observations across the full mass range to this analytical model as well as to the latest cosmological simulations we further conclude:

\begin{itemize}

    \item Our analytical model reproduces well the observed MZR assuming reasonable values for the gas depletion times, gas accretion rates, and the metallicity of infalling gas. However, we find that a $\sim2\times$ lower outflow efficiency ({\it i.e.} lower mass-loading factor) than found by the \textsc{EAGLE} and \textsc{FIRE} simulations is preferred to reproduce the MZR at $z=5$ (Figure~\ref{fig:mzrsim}).

    \item Our model suggests a fast metal enrichment. The observed metallicities at $z=5$ can be reached in less than $200\,{\rm Myrs}$. Furthermore, the range of metallicities {\it at a given stellar mass} is set by the stellar population age, the metallicity of inflowing gas, and outflow efficiency (mass-loading factor) (Figure~\ref{fig:mzrsim}). Due to the fast metal enrichment, metal-poor systems need either very efficient outflows ($>2\times$ more efficient that simulations predict) or pristine gas inflows.

    \item The observed scatter of the FMR at a give stellar mass or ($\mu$) can be explained naturally by a bursty star formation (Figure~\ref{fig:fmrscatter}). Less scatter is expected for more metal-rich galaxies, which can explain the $5\times$ lower scatter in the local FMR.

    \item Cosmological simulations generally reproduce the observed MZR, however, they tend to underpredict the metal enrichment of low-mass ($\rm \log(M/M_\odot)<8.5$) galaxies (Figure~\ref{fig:mzrcossim}). This may be because of the lack of simulated volume or dark matter halo resolution, or too slow enrichment of metals over time or too efficient outflows. 

    \item The \textsc{DUSTY-GAEA} simulation (which fits the observed MZR well at all stellar masses) suggests that the ALPINE-CRISTAL-JWST galaxies build up their metal abundance quickly at $z=15-10$, and then joined the observed $z=5$ FMR (Figure~\ref{fig:gaeasim}). The galaxies will then enrich further with metals to reach super-solar values at stellar masses of $\rm \log(M/M_\odot)>11.4$ ($5\times$ more massive than our Milky Way). They will leave the FMR relation at $z=2$ when star formation starts to get quenched.
\end{itemize}

Overall, galaxies at $z\sim5$ are already significantly metal enriched at all stellar masses. The difference in metal-enrichment at a given stellar mass can be explaind by the evolutionary age of the galaxies but also by differences in outflow efficiency and the metallicity of the accreted gas. The scatter in both the MZR and FMR can be reproduced by a stochastic SFH, but only at low masses at the onset of metal enrichment. Figure~\ref{fig:summaryfigure} summarizes these conclusions.

The outflow efficiency plays an important role in reproducing the observed MZR at $z\sim5$ as well as the observed \cplus~halos surrounding galaxies at the same epoch \citep[e.g.,][]{fujimoto19}. As discussed in our work and in \citet{pallottini25}, the outflows in simulations may be too efficient in removing metal-rich gas, causing an underestimation of the MZR. At the same time, lowering the outflow efficiency to fit the MZR may result in simulations not reproducing the extended \cplus~halos \citep[e.g.,][]{lunjunliu26,wang26}. Now, with more statistical data available, resolving this discrepancy may become an interesting topic in the near future.

\begin{acknowledgments}
{\small
This work is based in part on observations made with the NASA/ESA/CSA \textit{James Webb} Space Telescope.
The JWST and HST data presented in this article were obtained from the Mikulski Archive for Space Telescopes (MAST) at the Space Telescope Science Institute.
These observations are associated with programs, HST-GO-13641 (\dataset[doi: 10.17909/xne1-7v26]{https://doi.org/10.17909/xne1-7v26}), JWST-GO-01727 (\dataset[doi: 10.17909/ph8h-qf05]{https://doi.org/10.17909/ph8h-qf05}), JWST-GO-03045 (\dataset[doi: 10.17909/cqds-qc81]{https://doi.org/10.17909/cqds-qc81}), and JWST-GO-04265 (\dataset[doi: 10.17909/wac6-9741]{https://doi.org/10.17909/wac6-9741}).
This research has made use of the NASA/IPAC Infrared Science Archive (IRSA), which is funded by the National Aeronautics and Space Administration and operated by the California Institute of Technology. The following dataset was used from IRSA: \dataset[doi: 10.26131/IRSA178]{https://doi.org/10.26131/IRSA178}.
Support for program JWST-GO-03045 was provided by NASA through a grant from the Space Telescope Science Institute, which is operated by the Association of Universities for Research in Astronomy, Inc., under NASA contract NAS 5-03127.
This paper makes use of the following ALMA data: ADS/JAO.ALMA\#2017.1.00428.L, \#2019.1.00226.S, \#2022.1.01118.S, and \#2021.1.00280.L. ALMA is a partnership of ESO (representing its member states), NSF (USA) and NINS (Japan), together with NRC (Canada), MOST and ASIAA (Taiwan), and KASI (Republic of Korea), in cooperation with the Republic of Chile. The Joint ALMA Observatory is operated by ESO, AUI/NRAO and NAOJ.
This work made use of Astropy:\footnote{http://www.astropy.org} a community-developed core Python package and an ecosystem of tools and resources for astronomy \citep{astropy13, astropy18, astropy22}.
ALF thanks Jennifer Tan for proofreading the manuscript.
AF is partly supported by the ERC Advanced Grant INTERSTELLAR H2020/740120, and by grant NSF PHY-2309135 to the Kavli Institute for Theoretical Physics.
AN acknowledge support from the Narodowe Centrum Nauki (NCN), Poland, through the SONATA BIS grant UMO-2020/38/E/ST9/00077.
CG acknowledges support from the ASI-INAF agreement n. 2025-6-HH.0. 
DBS gratefully acknowledges support from NSF Grant 2407752.
DGE acknowledges support from ANID BECAS/DOCTORADO NACIONAL/2024-21251071.
EdC acknowledges support from the Australian Research Council through project DP240100589.
EI gratefully acknowledge financial support from ANID - MILENIO - NCN2024\_112 and ANID FONDECYT Regular 1221846.
H\"U acknowledges funding by the European Union (ERC APEX, 101164796). Views and opinions expressed are however those of the authors only and do not necessarily reflect those of the European Union or the European Research Council Executive Agency. Neither the European Union nor the granting authority can be held responsible for them.
IDL acknowledges funding from the European Research Council (ERC) under the European Union's Horizon 2020 research and innovation program DustOrigin (ERC-2019-StG-851622), from the Belgian Science Policy Office (BELSPO) through the PRODEX project ``JWST/MIRI Science exploitation'' (C4000142239) and from the Flemish Fund for Scientific Research (FWO-Vlaanderen) through the research project G0A1523N.
JM gratefully acknowledges support from ANID MILENIO NCN2024\_112.
JS is supported by JSPS KAKENHI (JP22H01262).
KVGC was supported by NASA through the STScI grants JWST-GO-04265 and JWST-GO-03777
LV acknowledges support from the INAF Minigrant ``RISE: Resolving the ISM and Star formation in the Epoch of Reionization'' (PI: Vallini, Ob. Fu. 1.05.24.07.01).
NGV acknowledges scholarship from ANID BECAS/Doctorado Nacional/2023-21231942
MA is supported by FONDECYT grant number 1252054, and  gratefully acknowledges support from ANID Basal Project FB210003 and ANID MILENIO NCN2024\_112.
MB gratefully acknowledges support from the ANID BASAL project FB210003. This work was supported by the French government through the France 2030 investment plan managed by the National Research Agency (ANR), as part of the Initiative of Excellence of Universit\'e C\^ote d'Azur under reference number ANR-15-IDEX-01.
MR acknowledges support from project PID2023-150178NB-I00 financed by MCIU/AEI/10.13039/501100011033, and by FEDER, UE. 
RA acknowledges financial support from projects PID2023-147386NB-I00 funded by MCIN/AEI/10.13039/501100011033 and by ERDF/EU and the Severo Ochoa grant CEX2021-001131-S to the IAA-CSIC.
RJA was supported by FONDECYT grant number 1231718 and by the ANID BASAL project FB210003.
SF acknowledges support from NASA through the NASA Hubble Fellowship grant HST-HF2-51505.001-A awarded by the Space Telescope Science Institute, which is operated by the Association of Universities for Research in Astronomy, Incorporated, under NASA contract NAS5-26555.
VV acknowledges support from the ANID BASAL project FB210003 and from ANID MILENIO NCN2024\_112.
}
\end{acknowledgments}

%

\vspace{5mm}
\facilities{HST(ACS), HST(WFC3/IR), JWST(NIRCam), JWST(NIRSpec), ALMA}


\software{\texttt{Astropy} \citep{astropy13,astropy18},
          \texttt{PyNeb} \citep{luridiana15}
          }



\clearpage
\newpage
\appendix
\setcounter{figure}{0}
\renewcommand\thefigure{\thesection.\arabic{figure}}  
\setcounter{table}{0}
\renewcommand{\thetable}{\thesection.\arabic{table}}

\section{Details on Metallicity Calculations}\label{app:metal}

\begin{figure}[b!]
\centering
\includegraphics[angle=0,width=1.0\columnwidth]{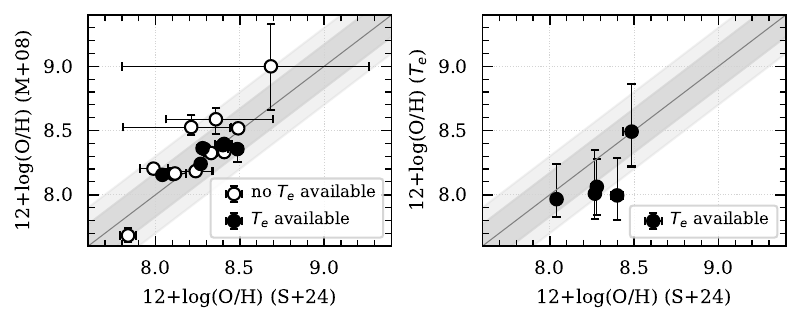}\vspace{-3mm}
\caption{
{\it Left:} Comparison of metallicity calculations based on the strong-line parameterizations from \citet{maiolino08} \citep[and][]{nagao06} and \citet{sanders24}. Filled symbols indicate the five galaxies with robust electron temperature measurements.
{\it Right:} Comparison of metallicity calculations based on the strong-line method \citep[using calibration by][]{sanders24} and the direct $T_e$ method measured from these data. This measurement is only available for five galaxies (filled symbols).
In both panels, the light and dark gray bands show deviations of $50\%$ and a factor of two ($\sim0.3\,{\rm dex}$), respectively.
\label{fig:metcomparison}}
\end{figure}

The left panel of Figure~\ref{fig:metcomparison} compares oxygen abundances based on the calibration from \citet{sanders24} with the one of \citet{maiolino08} \citep[see also][]{nagao06}. We do not find significant differences between these calibrations within the uncertainties of the measurements. The same holds for the calibration by \citet{hirschmann23}, which is similar to the one from \citet{sanders24} that is used here.

In some cases, we are able to compare the strong line metallicity measurements to the oxygen abundances independently derived from fainter optical auroral lines \citep[the ``$T_e$ method'', see review on metallicity measurements in ][]{kewley19}. These lines are only detected in a handful of galaxies; however, their detections allows an independent comparison to the metal abundances based on the strong line method.
We derived the oxygen abundances using the {\it getTemDen} module of the \texttt{PyNeb} Python package \citep{luridiana15} from the electron temperatures (\Toii~and \Toiii) and the electron density ($n_e$) by following these steps.

\begin{figure*}[t!]
\centering
\includegraphics[angle=0,width=\textwidth]{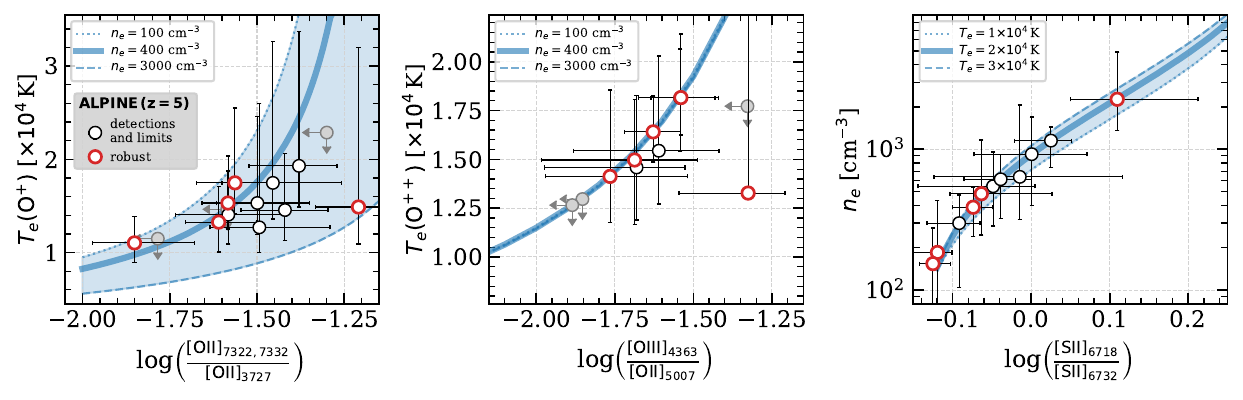}\vspace{-3mm}
\caption{Relation between line ratios, electron temperatures, and electron density derived from \texttt{PyNeb} (blue) including the measurements of the ALPINE-CRISTAL-JWST targets (symbols). The five galaxies indicated in red have the most robust $T_e$-based metallicity measurements (see also Figure~\ref{fig:metcomparison}).
\label{fig:temeasurement}}
\end{figure*}

\begin{enumerate}
    \item We derived \Toiii~from the [\ion{O}{3}]$_{4363}$/[\ion{O}{3}]$_{5007}$ line ratio. For this we assumed $n_e = 400\,{\rm cm^{-3}}$, however, the exact value is not important as this measurement is independent of $n_e$ (middle panel, Figure~\ref{fig:temeasurement}).

    \item We derived $n_e$ using the \SR~line ratio, assuming an electron temperature of $2\times10^4\,{\rm K}$ (again, this assumption is not critical as the dependence on electron temperature is weak, see right panel of Figure~\ref{fig:temeasurement}).

    \item We used the derived electron density to compute \Toii~using the line ratio ([\ion{O}{2}]$_{7322}$+[\ion{O}{2}]$_{7332}$)/[\ion{O}{2}]$_{3727}$ (left panel, Figure~\ref{fig:temeasurement}).

    \item Finally, using the {\it getIonAbundance} module of \texttt{PyNeb}, we computed the oxygen abundances $\frac{\rm O^+}{\rm H}$ and $\frac{\rm O^{++}}{\rm H}$ from the \Rtwo~and \Rthree~line ratios and the above measured electron temperatures (\Toii~and \Toiii) and electron density $n_e$.
\end{enumerate}

Note that we neglect higher ionization species in this calculation as they are not expected to significantly contribute to the total abundance.

In summary, we find electron temperatures of $1-2\times 10^4\,{\rm K}$ and electron densities of $100-1000\,{\rm cm^{-3}}$ ($\left<n_e\right>\sim 400\,{\rm cm^{-3}}$). These values are consistent with other measurements performed by JWST on galaxies of similar mass, SFR, and redshift \citep[e.g. from the {\it AURORA} survey;][]{topping25}.
The uncertainty in these parameters and the derived metal abundances are substantial due to the low signal-to-noise ratio of the auroral lines. We therefore only considered galaxies to have robust $T_e$-based metallicity measurements if they have a robust ($<50\%$ uncertainty) detection of the [\ion{O}{3}]$_{4363}$, [\ion{O}{2}]$_{7322}$, and [\ion{O}{2}]$_{7332}$ lines. The five galaxies that pass this criteria are indicated in red in Figure~\ref{fig:temeasurement}; {\it DC-417567}, {\it DC-630594}, {\it DC-709575}, {\it DC-842313}, and {\it DC-873321}.
The right panel of Figure~\ref{fig:metcomparison} shows a comparison of the $T_e$ and strong-line based metallicity measurements for these five galaxies. We find a good agreement between these independent measurements within the uncertainties, suggesting that our strong-line metal abundance estimates are reasonable.

\section{Details on Mass-Loading Factor}\label{app:massloading}

The outflow properties, specifically the mass-loading factor ($\epsilon_{\rm out}$), is closely linked to the evolution of galaxies. In our model, the outflow of metal-rich gas significantly affects the overall predicted metallicity. Importantly, the mass-dependence of $\epsilon_{\rm out}$ relates to the steepness of the MZR. 

Here, we use the characterization of $\epsilon_{\rm out}(M,z)$ as found in the \textsc{EAGLE} simulation \citep{schaye15,crain15}, specifically as presented in figure~3 of \citet{mitchell20}. 
To be able to use this result in our model, we derived an analytical representation in the form of a double power-law as follows:

\begin{equation}\label{eq:massloading}
    \frac{\left<\dot{M}_{\rm wind,ISM}\right>}{\left<\dot{M}_{\star}\right>} \equiv \epsilon_{\rm out} = \xi^{-1}\left(M_\star,z\right),
\end{equation}

with

\begin{equation} \label{eq:powerlaw}
    \xi\left(M_\star,z\right) = \frac{{\phi^*}(z)}{10^{(\delta+1)\times(\log(M_star) - \mathcal{M})} + 10^{(\gamma+1)\times(\log(M_star) - \mathcal{M})}},
\end{equation}

where we define $\phi^*(z) = 10^{1.1 - (1+z)^{0.09}}$,
$\mathcal{M}(z) = 8.2 + (1+z)^{0.29}$,
$\delta(z) = -0.4 - (1+z)^{0.02}$, and
$\gamma(z) = 0.05 - (1+z)^{0.01}$.

We find that this parameterization gives consistent results to \citet{muratov15}, which is based on the \textsc{FIRE} simulation \citep{hopkins14}.

\begin{figure}[t!]
\centering
\includegraphics[angle=0,width=0.8\columnwidth]{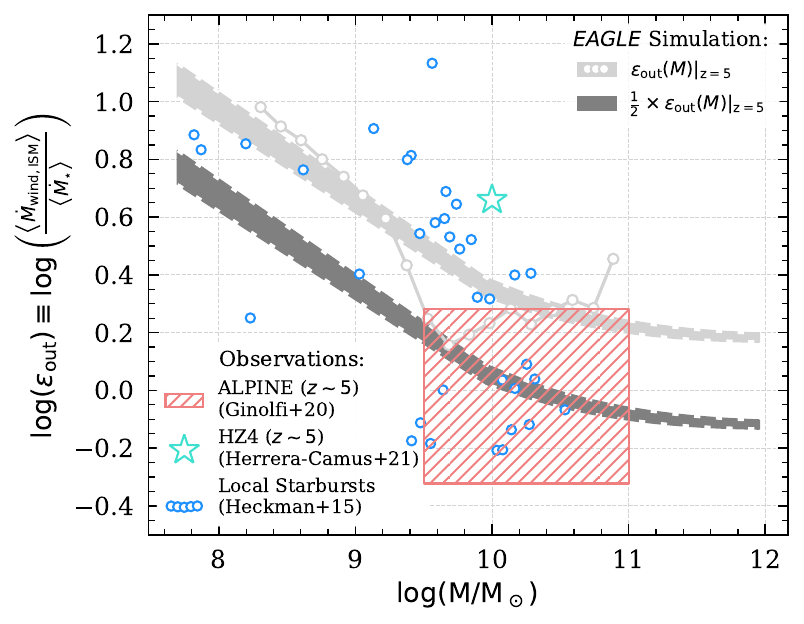}\vspace{-3mm}
\caption{Parameterization of the mass-loading factor $\epsilon_{\rm out}$ (Equation~\ref{eq:massloading}) derived by the \textsc{EAGLE} simulation \citep[figure 3;][]{mitchell20}. The light gray curve show the nominal outflow efficiency from \textsc{EAGLE}. The dark gray curve shows a two-times reduced outflow efficiency.
Also shown are results from the \textsc{ASTRAEUS} \citep[green dashed line;][]{ucci2023} and \textsc{FIRE-2} \citep[orange line;][]{Pandya2021}.
We compare this to observations at $z\sim5$ of the ALPINE sample \citep[red hatched square;][]{ginolfi20}, CRISTAL \citep[purple diamond;][]{birkin25}, and {\it HZ4} \citep[turquoise star;][]{herreracamus21}, as well as local starbursts \citep[blue circles;][]{heckman15} and a $z\sim8.5$ galaxy \citep[pink hexagon;][]{Algera2025}. Note that the ALPINE sample is more consistent with a reduced outflow efficiency, which is also preferred to reproduce the MZR (Figure~\ref{fig:mzrsim}).
\label{fig:epsout}
}
\end{figure}

Figure~\ref{fig:epsout} compares this theoretical mass-loading factor to observations.
The red hatched region shows the approximate area of the observed mass-loading factors of the ALPINE galaxies \citep{ginolfi20,birkin25}. The blue circles show the location of local starburst from \citet{heckman15} and the turquoise star denotes the measurement for {\it HZ4} at $z\sim5$ from \citet{herreracamus21}.
Most of the local starbursts as well as {\it HZ4} are consistent with the nominal mass-loading factor derived from the \textsc{EAGLE} simulation. However, the ALPINE galaxies are more consistent with an outflow efficiency reduced by a factor of two. Interestingly, such a reduction of the outflow efficiency is also preferred to reproduce the MZR of $z\sim5$ galaxies (Figure~\ref{fig:mzrsim}). This has also been noticed in \citet{pallottini25} and \citet{carniani24}. We note that the simulated mass-loading factors may be inferred differently compared to the observations which could cause this discrepancy \citep[e.g.,][]{pizzati20,carniani24}.

\bibliography{bibli}{}
\bibliographystyle{aasjournalv7}



\allauthors
\end{document}

%% file: table2.tex
\begin{table*}[ht!]
\centering
\setlength{\tabcolsep}{3pt}
\caption{Summary of physical properties of the ALPINE-CRISTAL-JWST galaxy sample.}
\label{tab:physical}
\begin{tabular}{cccccccccc}
\hline \hline
Name & $z_{\rm opt}$ & $\rm M_*$$^a$ & SFR$^b$ & Age$^c$ & \multicolumn{2}{c}{$\rm 12+log(O/H)$}$^d$ & $\rm T(O^{+})$ & $\rm T(O^{++})$ & $n_e$\\ \cline{6-7}
 & & $(\rm 10^9\,M_\odot)$ & $(\rm M_\odot\,yr^{-1})$ & (Myr) & (strong line) & ($T_e$--based) & ($\times 10^4\,{\rm K}$) & ($\times 10^4\,{\rm K}$) & ($\rm cm^{-3}$)  \\ \hline
DC-417567 & 5.666 & $10.0^{+0.3}_{-0.3}$ & $72^{+42}_{-27}$ & $105^{+100}_{-56}$ & $8.04^{+0.02}_{-0.02}$ & $7.96^{+0.28}_{-0.14}$ & $15^{+17}_{- 4}$ & $16^{+ 2}_{- 2}$ & $2268^{+2636}_{-912}$ \\
DC-494763 & 5.233 & $9.5^{+0.4}_{-0.4}$ & $28^{+36}_{-16}$ & $175^{+352}_{-119}$ & $8.29^{+0.02}_{-0.02}$ & -- & -- & $15^{+ 5}_{- 3}$ & -- \\
DC-519281 & 5.575 & $9.8^{+0.4}_{-0.4}$ & $32^{+35}_{-17}$ & $324^{+331}_{-194}$ & $7.99^{+0.08}_{-0.08}$ & -- & -- & -- & $637^{+1430}_{-321}$ \\
DC-536534 & 5.689 & $10.4^{+0.3}_{-0.3}$ & $62^{+64}_{-31}$ & $456^{+275}_{-224}$ & $8.10^{+0.04}_{-0.05}$ & -- & $14^{+ 6}_{- 3}$ & $18^{+ 2}_{- 3}$ & -- \\
DC-630594 & 4.440 & $9.7^{+0.3}_{-0.3}$ & $37^{+39}_{-19}$ & $186^{+227}_{-88}$ & $8.40^{+0.04}_{-0.04}$ & $7.99^{+0.29}_{-0.19}$ & $17^{+ 8}_{- 3}$ & $15^{+ 3}_{- 3}$ & $185^{+246}_{-116}$ \\
DC-683613 & 5.541 & $10.2^{+0.3}_{-0.3}$ & $68^{+67}_{-34}$ & $274^{+271}_{-159}$ & $8.49^{+0.03}_{-0.05}$ & -- & $19^{+14}_{- 4}$ & -- & $547^{+411}_{-262}$ \\
DC-709575 & 4.412 & $9.5^{+0.4}_{-0.4}$ & $28^{+39}_{-16}$ & $190^{+302}_{-95}$ & $8.27^{+0.02}_{-0.03}$ & $8.01^{+0.34}_{-0.20}$ & $13^{+ 4}_{- 3}$ & $18^{+ 3}_{- 2}$ & $484^{+685}_{-238}$ \\
DC-742174 & 5.636 & $9.5^{+0.4}_{-0.4}$ & $15^{+16}_{- 8}$ & $354^{+338}_{-215}$ & $7.84^{+0.05}_{-0.05}$ & -- & -- & -- & -- \\
DC-842313 & 4.552 & $10.6^{+0.5}_{-0.5}$ & $204^{+918}_{-167}$ & $394^{+250}_{-221}$ & $8.48^{+0.03}_{-0.05}$ & $8.49^{+0.37}_{-0.27}$ & $11^{+ 3}_{- 2}$ & $13^{+ 9}_{- 0}$ & $154^{+122}_{-87}$ \\
DC-848185 & 5.294 & $10.3^{+0.3}_{-0.3}$ & $178^{+290}_{-110}$ & $98^{+52}_{-48}$ & $8.28^{+0.02}_{-0.03}$ & $8.06^{+0.22}_{-0.22}$ & $15^{+ 3}_{- 3}$ & $14^{+ 4}_{- 2}$ & $388^{+153}_{-147}$ \\
DC-873321 & 5.154 & $10.0^{+0.3}_{-0.3}$ & $78^{+64}_{-35}$ & $112^{+84}_{-63}$ & $8.11^{+0.06}_{-0.07}$ & -- & $15^{+ 9}_{- 3}$ & -- & $613^{+307}_{-291}$ \\
DC-873756 & 4.545 & $10.5^{+0.1}_{-0.1}$ & $115^{+76}_{-46}$ & $550^{+317}_{-169}$ & $8.68^{+0.58}_{-0.88}$ & -- & -- & -- & -- \\
VC-5100541407 & 4.563 & $10.1^{+0.3}_{-0.3}$ & $42^{+50}_{-23}$ & $400^{+427}_{-239}$ & $8.21^{+0.24}_{-0.40}$ & -- & -- & -- & -- \\
VC-5100822662 & 4.520 & $10.2^{+0.3}_{-0.3}$ & $78^{+48}_{-30}$ & $237^{+196}_{-122}$ & $8.33^{+0.04}_{-0.04}$ & -- & -- & $15^{+ 4}_{- 3}$ & $299^{+174}_{-194}$ \\
VC-5100994794 & 4.580 & $9.6^{+0.3}_{-0.3}$ & $32^{+51}_{-20}$ & $193^{+252}_{-95}$ & $8.41^{+0.04}_{-0.05}$ & -- & $15^{+ 6}_{- 3}$ & -- & $1153^{+289}_{-411}$ \\
VC-5101218326 & 4.573 & $10.9^{+0.3}_{-0.3}$ & $562^{+534}_{-274}$ & $646^{+97}_{-275}$ & $8.36^{+0.34}_{-0.30}$ & -- & -- & -- & -- \\
VC-5101244930 & 4.581 & $9.7^{+0.3}_{-0.3}$ & $28^{+20}_{-12}$ & $183^{+365}_{-88}$ & $8.24^{+0.10}_{-0.13}$ & -- & $17^{+15}_{- 4}$ & -- & -- \\
VC-5110377875 & 4.550 & $10.2^{+0.2}_{-0.2}$ & $99^{+75}_{-43}$ & $161^{+127}_{-108}$ & $8.41^{+0.08}_{-0.09}$ & -- & $13^{+13}_{- 3}$ & -- & $924^{+767}_{-525}$ \\ \hline
\end{tabular}
\tablecomments{
$^a$ Stellar masses are derived in \citet{mitsuhashi24}.\\
$^b$ SFRs represent the total UV$+$FIR luminosities derived in \citet{faisst26}.\\
$^c$ Ages (light-weighted) are derived in \citet{faisst20b}.\\
$^d$ Metal abundance uncertainties include uncertainties from calibrations and line flux measurements, derived from extensive Monte-Carlo samplings. However, uncertainties due the dust attenuation corrections are not included.}
\end{table*}



%% file: table1.tex
\begin{table}
\centering
\caption{Fit to mass--metallicity relation (MZR; Figure~\ref{fig:mzrobs}) using all current data at $z=4-6$. Functional form:\\
$12+\log({\rm O/H}) = B \, \left( \log({\rm M_*/M_\odot}) - \log(M_0) \right)^2 + A$}
\label{tab:mzrfit}
\begin{tabular}{ccc} 
\hline \hline
$\log(M_0)$ & A  &  B  \\  \hline
$15.25\pm7.41$ & $8.82\pm0.97$ & $-0.02\pm0.02$ \\ \hline
\end{tabular}
\tablecomments{\raggedright Data included from: This work \citep[ALPINE-CRISTAL-JWST;][]{faisst26}; \citet{steidel14,sanders15,nakajima23,morishita24,curti23,scholtz25,sanders24,stanton24,sanders25,sarkar25,rowland25,stanton26}
}
\end{table}

%% file: table3.tex
\begin{table}
\centering
\caption{Scatter of the FMR measured at a given $\mu$ (and $\mu_0$) at different redshifts.
\label{tab:fmrscatter}}
\begin{tabular}{ccccc} 
\hline \hline
$z$ & $\sigma_{\rm FMR}$  &  $\hat{\sigma}_{\rm FMR}$$^\dagger$ & $\mu_0$ & Reference  \\ 
  & (dex)  &  (dex) &  &   \\ \hline
$z\sim0$ & $0.02$ & $<0.02$ & $0.22$ & (1) \\
 & $0.03$ & $<0.03$ & $0.55$ & (2) \\
$z\sim2.3$ & $0.16$ & $0.06$ & $0.60$ & (3) \\
$z\sim3.3$ & $0.22$ & $0.06$ & $0.60$ & (3) \\
$z\sim5$ & $0.24$ & $0.13$ & $0.66$$^\star$ & This work \\
 & $0.22$ & $0.10$ & $0.22$ & This work \\\hline
\end{tabular}
\tablecomments{
References: (1) \citet{mannucci10}; (2) \citet{curti20}; (3) \citet{sanders21}. \\
$^\dagger$ FMR scatter deconvolved by measurement uncertainties.\\
$^\star$ Consistent with $\mu_0=0.65$ obtained by \citet{curti24}.
}
\end{table}